\documentclass[12pt]{article}
\usepackage[colorlinks=true,linkcolor=blue,citecolor=blue,urlcolor=blue]{
hyperref}

\usepackage{scicite}
\usepackage{times}
\usepackage{amsmath,amssymb,soul,color}
\usepackage{graphicx}
\usepackage{fancyhdr}
\usepackage{ifoddpage}

\newcommand{\oddPageText}{Transcriptome derived interaction networks}

\newenvironment{sciabstract}{%
\begin{quote}}
{\end{quote}}

\newcommand{\keywords}[1]
{
  \small
  \textbf{Keywords:} #1
}

\title{Inferring interaction networks from transcriptomic data: methods and
applications}

\author
{Vikram Singh$^{\dagger}$, Vikram Singh$^{\ast}$\\
\\
\normalsize{Centre for Computational Biology and Bioinformatics, Central
University of Himachal Pradesh,}\\
\normalsize{Dharamshala, 176206, Himachal Pradesh, India}\\
\\
\normalsize{$^\ast$E-mail:  vikramsingh@cuhimachal.ac.in}
}

\date{}

\begin{document} 


\baselineskip24pt


\maketitle


\begin{sciabstract} \bf
Transcriptomic data is a treasure-trove in modern
molecular biology, as it offers a comprehensive viewpoint into the
intricate nuances of gene expression dynamics underlying biological systems.
This genetic information must be utilised to infer biomolecular interaction
networks that can provide insights into the complex regulatory mechanisms
underpinning the dynamic cellular processes. Gene regulatory networks and
protein-protein interaction networks are two major classes of such
networks. This chapter thoroughly investigates the wide range of methodologies
used for distilling insightful revelations from transcriptomic data that
include association based methods (based on correlation among
expression vectors), probabilistic models (using Bayesian and Gaussian models),
and interologous methods. We reviewed different approaches for evaluating the
significance of interactions based on the network topology and biological
functions of the interacting molecules, and
discuss various strategies for the identification of functional modules. The
chapter concludes with highlighting network based techniques of
prioritising key genes, outlining the centrality based, diffusion
based and subgraph based methods. The chapter provides a meticulous framework
for investigating transcriptomic
data to uncover assembly of complex molecular networks for their adaptable
analyses across a broad spectrum of biological domains.
\end{sciabstract}

\begin{sciabstract}
\end{sciabstract}

\section{Introduction}\label{sec:intro}

Rooted in the foundational discoveries of biological sciences is the central
dogma of molecular biology, encapsulating the principle that DNA makes RNA,
that further directs protein
synthesis\cite{crick1970central,bustamante2011revisiting}. Genetic information
coded in DNA is transcribed into RNA through a complex, precise, and
meticulously
directed cellular process called transcription. The regulatory and housekeeping
RNAs in concert with proteins
intricately engage in spatiotemporally regulated interactions, finely tuned by
specific environmental cues, to govern the multitude of life sustaining
processes \cite{jacob1961genetic,stark1978ribonuclease,morris2014rise}. In
recent decades, the abundance of genomic sequences has markedly expanded,
thanks to the successful completion of a number of genome projects and the
sophistication of next-generation sequencing technologies
\cite{siva20081000}. Nevertheless, bridging the gap
between gene expression data and the cellular function continues to stand as a
formidable challenge \cite{segal2002decomposing}. Extensive transcriptome-wide
studies
need to be undertaken in order to effectively bridge this gap and
determine the functional groupings among genetic elements. With the advent of
high-throughput technologies, including microarrays, RNA sequencing (RNA-seq),
and the corresponding development of sophisticated data analysis methods, our
ability to comprehend gene function and its relevance to diseases has undergone
a major transformation. These cutting-edge techniques provide a systematic view
of the functional status of genes, allowing us to see beyond the individual
elements of the genetic circuitry and appreciate the unified whole
\cite{lowe2017transcriptomics}.

Unravelling the molecular mechanisms of biological processes, and
understanding the optimal utilisation of intricate genetic circuitry as well as
their potential implications in diseases hinges on deciphering the complex
interaction patterns among these genetic elements \cite{emilsson2008genetics}.
Thus, one of the primary objectives in biological research is to systematically
identify all molecules
within a living cell and the relationship among them to explain the phenotypic
variability. Extensive research has demonstrated that the complex regulation
governing gene expression is shaped by cis-acting DNA elements and complex
protein interactions \cite{maloy1993autogenous,maniatis1987regulation}.
Moreover, an evolving understanding of gene regulation has also unveiled the
possibility
of trans regulation, highlighting the nuanced interplay between DNA and
transcribed RNA \cite{killary1984genetic}. However, the functions of many
genes are still less understood, a situation that has only become more
complex with the recent identification of many novel noncoding genes. Recently,
multiple genes were observed to be coexpressed, clustering together in patterns
of expression. An observation led to the hypothesis that coexpression may be
the result of the coregulation of genes, implicating complex regulatory
networks \cite{wen1998large}. These discoveries sparked the investigation of
broader networks underlying gene coexpression, with the potential to shed light
on the convoluted coordination of cellular processes. The coexpression networks
represent the cooperative behaviour of genes in response to different
conditions. Subsequently, the constructed networks become the subject of
in-depth examination and analyses, providing a comprehensive view of gene
interactions and their implications.

As we delve deeper into the domain of network construction using transcriptomic
data, we investigate how these networks reveal the symphony of molecular
interactions within cells and the complexities of biological organisation. In
Section \ref{sec:pot}, we embark on a brief overview of the various platforms
utilised for transcriptomic data generation followed by an overview of graph
theory in Section \ref{sec:bin}. Moving into Section \ref{sec:iin}, we  delve
into the core ideas associated with construction of interaction networks,
exploring three pivotal methods: association based methods, probabilistic
methods and interologous network construction techniques. In Section
\ref{sec:na}, we focus on analysing the constructed network. Here, we  uncover
the significance of the predicted interactions within these networks, a
critical step in deciphering their statistical and biological relevance.
Subsequently, we discuss the methods to identify functional modules within
these networks, elucidating the coordinated actions of genes or proteins
in specific cellular processes. Finally, we elaborate
various approaches to identify key drivers and molecular architects that govern
the dynamics over these networks and ultimately control the cellular response.

\section{A primer on transcriptomics}\label{sec:pot}

Almost all of the processes within a living organism, like, gene
expression, cellular regulation, and
disease modulation are substantially affected by the RNA molecules that
include protein-coding mRNAs and non-coding RNAs, such as transfer RNAs
(tRNAs), ribosomal RNAs
(rRNAs), microRNAs (miRNAs), and long non-coding RNAs (lncRNAs)
\cite{carninci2005transcriptional}. Thus, the
transcriptome encompasses the entirety of RNA molecules present in a given cell
or tissue in a specific temporal context \cite{jacquier2009complex}. This
compilation reveals a
multidimensional comprehension of cellular behaviour by highlighting the
complicated gene expression dynamics. Transcriptomics, a systematic
investigation of the transcriptome, provides unprecedented insights into the
genetic complexities that regulate cellular responses to developmental cues,
environmental fluctuations, and disease
manifestations\cite{jacquier2009complex}. Transcriptomics offers a panoramic
perspective of the complex cellular dynamics, as it provides a
comprehensive platform which transcends isolated gene analysis
and gives a holistic understanding of the molecular dynamics driving
biological systems. The rapid evolution of transcriptomics has been propelled
by technological advancements \cite{lowe2017transcriptomics}. Early attempts
at understanding individual
transcripts predated comprehensive transcriptomics approaches. Techniques like
expressed sequence tags (ESTs) \cite{marra1998expressed} and low-throughput
Sanger sequencing \cite{sanger1977dna} were utilised to sequence individual
transcripts, providing insight into gene
content. One of the pioneering sequencing-based methods, serial analysis of
gene expression (SAGE) \cite{velculescu1995serial}, emerged in 1995.
However, the comprehensive
expression and regulation of the entire transcriptome remained elusive until
high-throughput techniques such as microarrays and RNA-Seq emerged
\cite{lowe2017transcriptomics}.

\begin{figure}[h]
  \centering
  \includegraphics[width=0.7\textwidth]{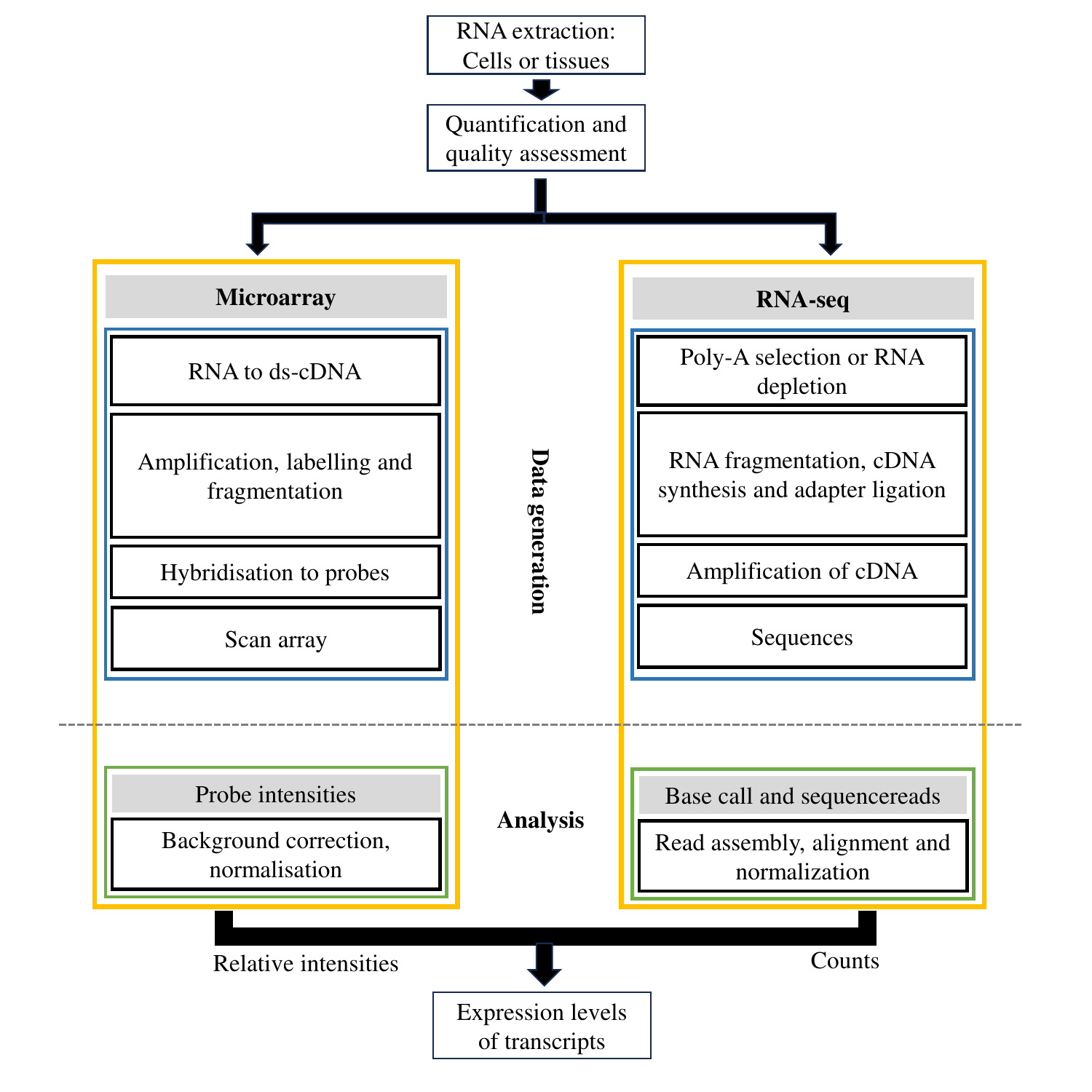}
  \caption{An overview of the general steps involved in generating and
preprocessing transcriptomic data using Microarray and RNA-seq technologies.}
  \label{fig:GA}
\end{figure}

Over the past few decades, two of the most commonly employed transcriptomics
technologies have been microarray \cite{schena1995quantitative} and
RNA-sequencing (RNA-seq) \cite{ozsolak2011rna}.
Microarrays utilise hybridisation, involving thousands or even millions of
short nucleotide oligomers known as probes strategically arrayed on a solid
substrate (Fig \ref{fig:GA}), typically glass. Fluorescently labelled
transcripts are hybridised
with these probes, resulting in fluorescence intensity at each probe location
reflecting the corresponding transcript’s abundance
\cite{schena1995quantitative}. Microarrays generally
fall into two main categories: low-density spotted arrays and high-density
short probe arrays. Low-density arrays involve minute droplets of purified
cDNAs placed on a glass slide. These arrays feature longer probes compared to
high-density arrays, impacting their transcript resolution. Fluorescence ratios
from different fluorophores are employed to calculate a relative measure of
abundance in spotted arrays \cite{shalon1996dna}. In contrast, high-density
arrays employ
single-channel detection, with each sample hybridised and detected individually
\cite{lockhart1996expression}. Pioneered by the Affymetrix GeneChip array,
high-density arrays utilise multiple short 25-mer probes to quantify individual
transcripts, collectively assessing one gene. Microarrays are relatively less
expensive and can be used to analyse a large number of genes simultaneously.
However, they are less sensitive than RNA-seq and challenging to interpret.
Another major drawback associated with microarrays is that these arrays
necessitate a prior understanding of the organism under investigation, often
derived from an annotated genome sequence or a library of expressed sequence
tags (ESTs) that serve as the basis for generating the probes
\cite{lowe2017transcriptomics}.

The advent of cutting-edge technologies, notably RNA sequencing, has brought a
revolutionary paradigm shift to transcriptomics \cite{wang2009rna}. This
innovation enables
the precise and comprehensive dissection of the RNA dynamics. RNA-Seq
encompasses several crucial steps (Fig \ref{fig:GA}) involving RNA isolation,
library preparation and sequencing. The process initiates with the extraction
of RNA molecules from a biological sample. This step is pivotal to ensure the
integrity and representation of the transcriptome. Subsequently, a cDNA library
is synthesised that includes RNA fragmentation, reverse transcription to
synthesise
complementary DNA (cDNA), adapter ligation and quality control using
quantitative rt-PCR. Once the library is ready, the sequencing
phase commences. High-throughput sequencing platforms decode the cDNA
fragments, generating millions of short sequences or reads
\cite{ozsolak2011rna}. Sequenced
nucleotide fragments are aligned to reference genomes or assembled de novo to
reconstruct original RNA transcripts. RNA-Seq offers a broad dynamic range,
enabling gene identification, real-time activity analysis, and accurate
modelling of gene expression levels. Ongoing improvements in DNA sequencing
technology have enhanced RNA-Seq’s throughput, accuracy, and read length,
leading to its widespread adoption over microarrays. RNA-Seq requires lower RNA
input than microarray, enabling finer examination of cellular structure at the
single-cell or nuclear levels
\cite{van2013rna,huber2015orchestrating,amezquita2020orchestrating}.

\section{Biomolecular interactions from graph theoretic
perspectives}\label{sec:bin}

\begin{figure}[h]
  \centering
  \includegraphics[width=0.7\textwidth]{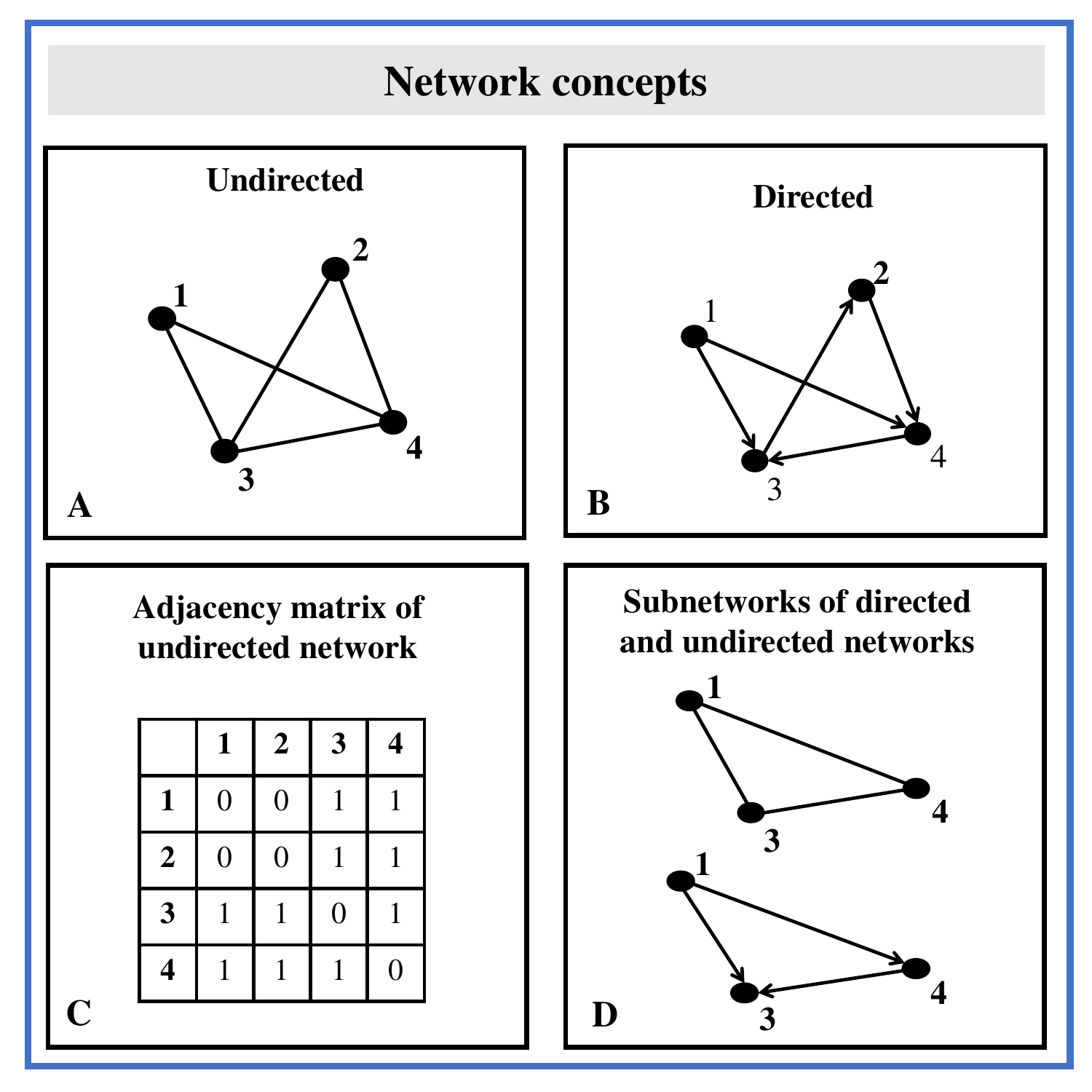}
  \caption{Depiction of fundamental network theoretic concepts. (A) introduces
an undirected network featuring four nodes and five edges, each node boasting a
degree of two.(B) A directed network comprising of four nodes and five edges.
Notably, nodes 3 and 4 exhibit an in-degree of two and an out-degree of one,
while node 2 possesses a balanced in-degree and out-degree of one, and node 1
showcases an out-degree of two alongside an in-degree of zero. (C) Adjacency
matrix corresponding to the aforementioned undirected network, offering a visual
representation of inter-node connections. (D) Subgraphs on nodes 1, 3, and 4
derived from the networks (a) and (b).}
  \label{fig:NC}
\end{figure}

The biological world is organised into a hierarchical structure,
spanning from the atomic and molecular level to ecosystems and the biosphere
\cite{macmahon1978levels}. At each tier of this biological hierarchy, and even
across different levels, interaction networks play crucial regulatory roles.
Networks serve as a fundamental framework for understanding complex
relationships within biological systems
\cite{newman2003structure,albert2002statistical}. Whether we are
examining molecular interactions within a cell, the flow of energy and nutrients
in ecosystems, or the intricate web of connections in the human brain, these
systems have been aptly represented and analysed using network frameworks
\cite{bascompte2009disentangling}. In this context, an interaction network
(Fig \ref{fig:NC}A, B) is essentially a graph \(G = (V, E)\) comprising of a
set of nodes \(V\) and a set of edges \(E \subset V \times V\). Graph theory
provides the formal foundation to explore and decipher the structural and
functional properties of these networks, making it an indispensable tool for
researchers. By treating biological entities as nodes and their interactions as
edges, we can employ the rich toolbox of graph theory to gain insights into the
organisation, dynamics, and emergent properties of these systems
\cite{newman2003structure}. These interactions underpin processes like signal
transduction, gene regulation, and metabolism. However, deciphering these
interactions, especially within the complex milieu of living cells, can be an
immensely challenging endeavour.
Fortunately, there exists an alternative approach, one that provides us with a
computational lens into the intricate world of molecular interactions. This
approach involves measuring the abundance of mRNA levels within cells. By
quantifying the expression of thousands of genes simultaneously, researchers can
generate detailed snapshots of gene expression patterns
\cite{gardner2003inferring}. It is within these expression patterns that a
wealth of information about underlying molecular interactions is concealed.
Using mRNA expression data, computational methods can be employed to reconstruct
the interaction structure (network topology) that governs these patterns
\cite{d2000genetic}.

Networks constructed from expression data would be directed (Fig
\ref{fig:NC}A) or undirected (Fig
\ref{fig:NC}B) graphs determined by the
statistical model employed \cite{wang2014review}. In graph
theory, a directed network, often referred to as a directed graph or digraph,
contains edges that are ordered pairs of vertices and thus have a specific
orientation or direction associated with them. These edges represent causal
relationships or information flow from one node (gene or protein) to another
\cite{friedman2000using}. Conversely, an undirected network consists of edges
that lack direction and typically represent symmetric associations, such as
co-expression relationships between genes or proteins
\cite{d2000genetic,horvath2008geometric}. The adjacency matrix \(\bf A\) is a
fundamental representation of networks, with rows and columns corresponding to
nodes and entries indicating the presence or absence of edges (Fig
\ref{fig:NC}C). Each entry \(a_{i,j}\) of
the \(\bf A\) is defined as
\[ a_{i,j} =
  \begin{cases}
    x_{i,j}       & \text{if } \{i, j \in E\}\\
    0       & \text{otherwise }
  \end{cases}
\]
where \(x_{i,j}\) would vary according to some weight function
\(w_{i,j}\) or have values \(1\) if the network is unweighted. The adjacency
matrix for undirected networks is symmetric, i.e. if an edge exists between
nodes  \(i \text{ and } j\), the corresponding edge between nodes  \(j \text{
and } i\) is also
present. In directed networks, the matrix can be asymmetric, reflecting the
direction of interactions \cite{estrada2012structure}. Some essential
topological properties of a graph include adjacency i.e. if nodes \(i, j \in E
(G)\), then node \(i\) is called adjacent to or neighbour of node \(j\). All the
nodes adjacent to node \(i\) constitute its neighbourhood \( N(i)\), which is
defined as \(N(i) = j \in V(G) | i,j \in E (G)\) \cite{assenov2008computing}.
The degree of a node \(i\), denoted as \(deg(i) = |N(i)|\), represents the total
number of edges incident to a node \(i\), offering insights into a node's
centrality. In directed networks, we distinguish in-degree (incoming edges) and
out-degree (outgoing edges), discerning the nature of interactions
\cite{assenov2008computing}. Similarly, several other properties describe the
general structure, node rank or
local interconnectivity patterns between network nodes. These properties include
the clustering coefficient, which mathematically quantifies how nodes tend to
cluster, highlighting localised structures. Eigenvalues \(\lambda\) and
eigenvectors unveil global network properties and central nodes, a key tool in
identifying influential network components. The graph spectrum, derived from
eigenvalues, encapsulates structural characteristics and stability
\cite{assenov2008computing}.

\clearpage

\section{Inferring interaction networks from RNA-seq data}\label{sec:iin}

\begin{figure}[h]
  \centering
  \includegraphics[width=0.7\textwidth]{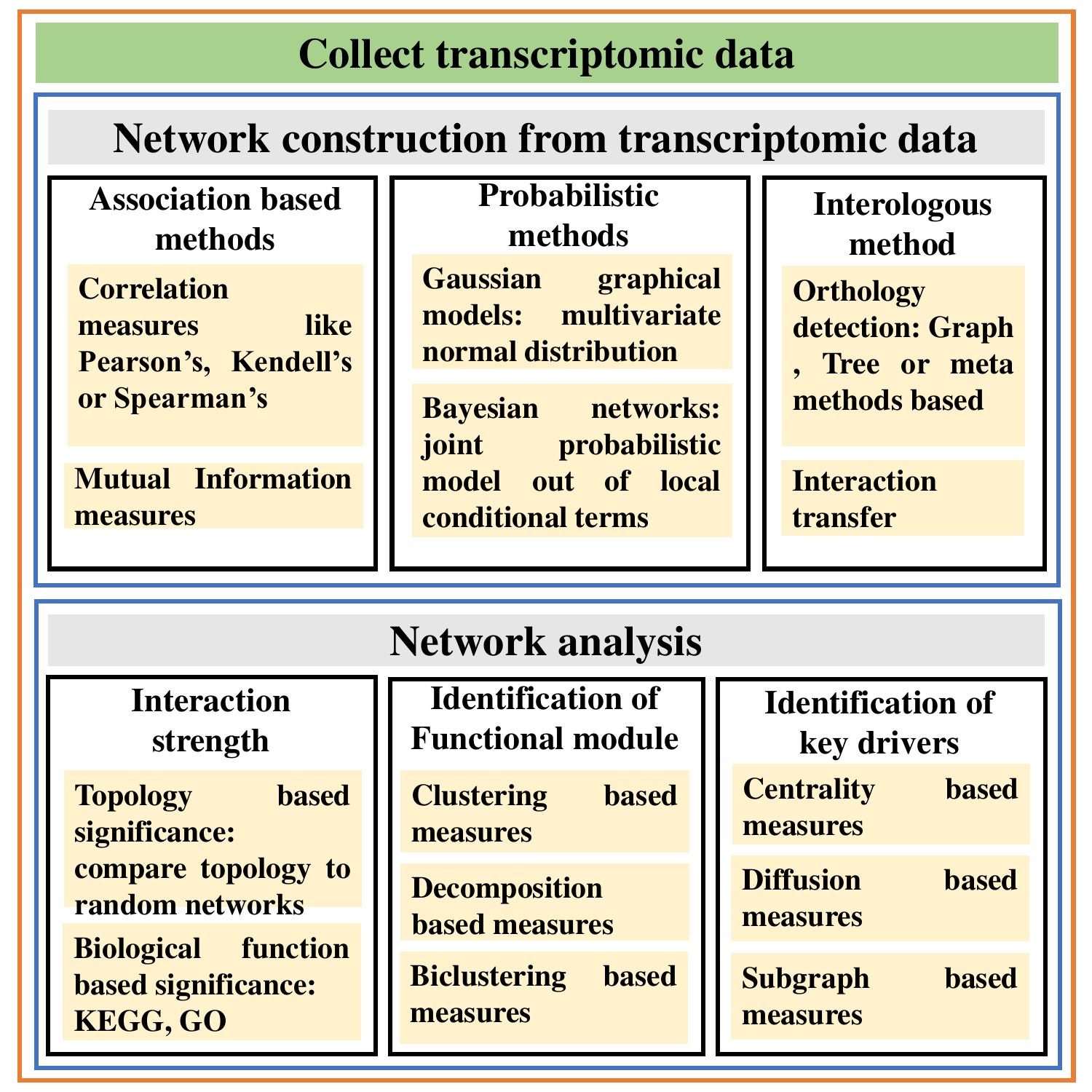}
  \caption{An overview of (A) various methods used to infer interaction networks
from transcriptomic data and (B) Different network analysis methods.}
  \label{fig:MO}
\end{figure}

Differentially expressed genes (DEG) analysis often yields a substantial
matrix, where each row represents a gene, and the columns represent attributes
such as fold change, \emph{p-values}, or other relevant measures
\cite{wang2010degseq,kvam2012comparison,slonim2002patterns}. At this
juncture, the sheer volume of data can be overwhelming for any scientist.
Extracting meaningful insights and interpreting the biological significance
necessitates a reduction in data complexity and the discovery of inherent
structures within it. To tackle this challenge, several standard and widely
applied methodologies come into play. These include clustering techniques
\cite{jain1988algorithms} and principal component analysis
\cite{raychaudhuri2000principal}, which help identify patterns and group
genes with similar expression profiles. For more intricate analyses,
researchers delve into inferring genome-scale networks, such as protein
interaction networks \cite{pellegrini2004protein}, or leverage knowledge-based
networks like pathway networks \cite{kanehisa2000kegg} and gene coexpression
networks \cite{van2018gene} derived from the expression
data. These advanced approaches offer a deeper understanding of the interplay
between genes and pathways, facilitating the extraction of biologically
relevant information from the transcriptomic data. In the following, we discuss
three different strategies (Fig \ref{fig:MO}) for biomolecular
interaction networks construction using transcriptomic data, including
association based networks, probabilistic networks and interologous networks.

\clearpage

\subsection{Association based interaction networks}\label{subsec:ain}

Network approaches have been employed in molecular biology to explore the
complex interaction patterns underlying expression data
\cite{pellegrini2004protein,kanehisa2000kegg,van2018gene}. The nodes in
these types of networks typically represent gene expression levels, and the
edges that connect them represent interactions, which can be formulated either
by correlation, mutual information or some other mathematical model used to
represent the system. It is crucial to remember that interactions in
association networks based on correlation and mutual information do not
identify the direction of effect between interacting genes and thus are
undirected. Interactions in probabilistic networks, on the other hand, are
directed. The initial step in building association networks is to create a
distance matrix from expression data, a fundamental process for network
analysis \cite{langfelder2008wgcna}. This step typically
involves calculating numeric values that quantify the differences between gene
expression patterns. Commonly, pairwise correlations or covariances between
genes of interest are computed using expression values. There
exists a number of correlation measures like Pearson's \cite{pearson1895vii},
Kendall's \cite{kendall1938new} and Spearman's \cite{spearman1961proof} that
have been commonly employed for this purpose
\cite{song2012comparison,kumari2012evaluation}.
Alternatively, one can use information theory or more specifically
mutual information score to measure the resemblance between pairs of expression
vectors \cite{butte1999mutual,song2012comparison}. The outcome of
this step is an \(m \times m\) matrix, with \(m\) representing the number of
analysed genes. Each element within this \(m \times m\) matrix signifies the
distance between a pair of elements and is commonly referred to as a distance
matrix, laying the groundwork for subsequent association network construction.

The distance matrix must then be converted to an adjacency matrix containing
only those elements which survive specific measures of significance
\cite{song2012comparison,lopez2022network}. Constructing an adjacency matrix
from the initial distance matrix is a pivotal step in the creation of
association networks. The purpose of this transformation is to identify
significant relationships or connections between elements, effectively
delineating the network’s edges \cite{song2012comparison}. The resulting
adjacency matrix is a binary representation, where surviving connections are
marked as ones, signifying edges and non-surviving entries are marked as zeros,
indicating no connection. Converting the distance matrix into an adjacency
matrix is of paramount importance as it profoundly influences the structure and
interpretability of the network. Several approaches can be employed for this
conversion, however, the most commonly employed approach is the threshold-based
method \cite{lopez2022network}. Thresholding involves defining a threshold value
and categorising the values in the distance matrix based on whether they meet
or exceed this threshold. Two primary types of thresholding methods are
used in this context: hard and soft.

Hard thresholding is a straightforward method for converting a distance matrix
into an adja- cency matrix, a critical step in coexpression network
construction \cite{paci2017swim}. The process involves setting a predetermined
threshold value, often determined through heuristics or statistical
considerations. All values in the distance matrix below this threshold are
assigned a value of zero, indicating insignificance, while those equal to or
above the threshold are retained as connections in the adjacency matrix
\cite{butte2000determining,carter2004gene}. One of the notable
advantages of hard thresholding is its simplicity. It is easy to implement and
interpret, providing a clear binary distinction between significant and
insignificant connections. Moreover, it is computationally efficient, as it
involves simple comparisons and assignments. However, this simplicity comes at
a cost. Hard thresholding can lead to the loss of valuable information
regarding the strength of connections below the threshold, potentially
disregarding subtle but biologically relevant relationships. Further-
more, selecting an appropriate threshold value can be challenging, and the
method is sensitive to this choice, with slight variations in threshold leading
to significantly different network structures \cite{carter2004gene}.

Soft thresholding in contrast, is a more sophisticated approach to converting a
distance matrix into an adjacency matrix. This method begins by computing
correlation coefficients between elements in the distance matrix. These
coefficients are then subjected to a power transformation, typically through
squaring or applying other functions. This transformation accentuates stronger
correlations while attenuating weaker ones. Subsequently, a hard threshold
is applied to these transformed values, resulting in a weighted adjacency
matrix \cite{langfelder2008wgcna}. In this weighted matrix, values below the
threshold contribute to the network but with reduced weight, preserving
information about the strength of relationships. One of the primary advantages
of soft thresholding is its ability to retain information about the strength of
connections, allowing for a more nuanced representation of the network. This
feature can help
reduce noise and capture subtle coexpression patterns that might be lost in a
binary representation. It also provides flexibility in controlling the
network’s sparsity, as researchers can adjust the threshold to balance the
number of retained connections \cite{zhang2005general}. However, this method
introduces complexity
due to the mathematical transformations involved, potentially making it more
computationally intensive. Additionally, the selection of the power
transformation exponent and threshold can be subjective, although they provide
researchers with more options for fine-tuning the network’s characteristics.
The interpretation of weighted networks can also be more
intricate than binary networks, often requiring additional analytical
techniques. The final unweighted or weighted adjacency matrix thus obtained
represents the network of interest, which is further analysed using graph
theoretic approaches to explore coexpression patterns among different genes
under different conditions.

\subsection{Probabilistic networks}\label{subsec:pn}

Previously described data-driven methods rely on correlation or
information-theoretic measures of dependence but do not explicitly define a
probabilistic data model. In this section, we discuss two distinct classes of
methods that explicitly begin with a probabilistic data model
\cite{wang2014review}. These methods
employ global measures of fit, such as joint likelihood, or employ Bayesian
approaches to discern the underlying network structure.

\subsubsection{Gaussian graphical models}\label{subsubsec:ggn}
Gaussian graphical models (GGMs) are a powerful tool in statistical modelling
that uses a graph to model the conditional dependence structure among continuous
random variables \cite{schafer2005empirical}. GGMs are based on the assumption
that the variables are jointly Gaussian, meaning their distribution is a
multivariate Gaussian distribution. The graph consists of nodes representing
random variables (gene expression measurements in our case) and edges
representing conditional dependencies. Two variables are conditionally
independent given a set of other variables in the model if their joint
distribution is the same as the product of their conditional distributions. If
an edge does not connect two nodes, they are conditionally independent, given
all the other variables in the model. The graph can be either directed or
undirected, depending on the type of conditional dependencies. The joint
distribution of the random variables in a Gaussian graphical model is a
multivariate normal distribution, which is given
by \[ p({\bf x|\mu, \Sigma}) = \frac{1}{2\pi^{\frac{n}{2}} |{\bf
\Sigma}|^{\frac{1}{2}}} \text{ exp }
\left[-\frac{1}{2} ({\bf x - \mu})^T {\bf \Sigma}^{-1} ({\bf x - \mu}) \right]\]
Where \({\bf x \in \mathbb{R}^n}\) is a multivariate normal vector, \({\bf
\mu}\) is the mean vector, and \({\bf \Sigma}\) is the covariance matrix.
The inverse of the covariance matrix, \({\bf \Omega = \Sigma}^{-1}\), is called
the precision matrix and it encodes the partial correlations between the random
variables, which measures the extent of correlation between these two
variables after adjusting for the
effect of the other variables. The precision matrix has a zero entry if and
only if the corresponding nodes are conditionally independent in the graph.
This means that the graph can be obtained from the sparsity pattern of the
precision matrix, and vice versa.
The main problem in Gaussian graphical models is to estimate the precision
matrix from data \cite{edwards2012introduction}, which can be done by maximising
the log-likelihood function:

\[\ell(\mathbf{\Sigma}^{-1}) = \log |\mathbf{\Sigma}^{-1}| -
\mathrm{tr}(\mathbf{S}\mathbf{\Sigma}^{-1}) - n\log(2\pi)\].
The term \(- n\log(2\pi)\) is a constant that does not depend on
\(\mathbf{\Sigma}^{-1})\), and it is added to make the log-likelihood function
consistent with the probability density function of the multivariate normal
distribution. The term \(\log |\mathbf{\Sigma}^{-1}|\) is the natural logarithm
of the determinant of the precision matrix, which measures the information
content of this matrix. The term
\(\mathrm{tr}(\mathbf{S}\mathbf{\Sigma}^{-1})\) is the trace of the product of
the sample covariance matrix and the precision matrix, which measures how well
the precision matrix fits the data. This is a convex optimisation problem
that can be solved by various methods, such as gradient descent, Newton's
method, or coordinate descent. However, when the number of variables is large
compared to the number of observations, the problem becomes ill-posed and the
maximum likelihood estimator does not exist or is not unique. In this case,
regularisation techniques are needed to impose some constraints on the
precision matrix, such as sparsity or low rank. Some common regularisation
methods include lasso that adds an \(L1\) penalty on the entries of
the precision matrix, which encourages sparsity and selects relevant edges in
the graph. Graphical lasso applies lasso on each row (or column) of the
precision matrix separately, which ensures positive definiteness and symmetry.
Sparse inverse covariance estimation (SICE) applies graphical lasso on a
penalised log-likelihood function that incorporates prior information on the
graph structure \cite{friedman2008sparse}.

\subsubsection{Bayesian interaction networks}\label{subsubsec:bin}

The methods discussed above range from analysing marginal dependencies
in coexpression measures to exploring conditional dependencies in partial
correlation approaches, and strive to unveil gene relationships based on
various probabilistic dependencies. However, it is essential to note that these
methods, consequently, cannot represent causal relationships between genes. In
contrast, Bayesian networks provide a framework for modelling and studying the
causal dependencies among genes \cite{pearl1988probabilistic}. It consists of
nodes, which represent variables, and edges, which represent conditional
dependencies \cite{needham2007primer}. A Bayesian network can be used to model
gene expression data by assuming that each gene is a variable whose expression
level depends on the expression levels of some other genes
\cite{friedman2000using}. To generate a Bayesian network from gene expression
data \(D\), one needs to find the optimal network structure \(G\) that best fits
the data, i.e., the set of nodes and edges that represent the actual gene
interactions. The structure is a directed acyclic graph (DAG) where each node
represents a variable, and each edge represents a conditional dependence. It
represents the joint probability distribution of the genes as a product of local
conditional probability distributions, each depending on the state of the parent
nodes in the graph.

\[P(X_1, \ldots, X_p) = \prod_{i = 1}^{p} P(X_i|Pa^{G} (X_i))\]

where \(Pa^{G}\) are all the parent nodes of \(X_i\) in the DAG \(G\). This
joint
probability distribution represents a set of conditional independence
relationships among the variables, meaning that each variable is independent of
its non-descendants, given its parents. It is called the Markov assumption
that, together with a restricted number of parents for each node, facilitates
the definition of conditional probabilities and the effective execution of
inference in a Bayesian network. Every directed acyclic graph \(G\) is assessed
using a Bayesian score, represented as the posterior probability of \(G\) that
is
given by
\[S (G: D) = log P(G|D) =  log P(D|G) + log P(G) + C\].

Computing posterior probabilities in DAG modeling
involves two key steps: learning the DAG structure from observational data and
estimating conditional probabilities given this structure. This task is
challenging due to the super-exponential growth in possible DAG topologies and
the high dimensionality of expression data, leading to many equally scoring
DAGs. To tackle this complexity, heuristic algorithms like greedy hill-climbing,
simulated annealing, and genetic algorithms \cite{yu2002using} are employed,
iteratively adjusting the DAG topology. Biologically informed constraints and
coexpression clustering help reduce the search space. Instead of selecting a
single optimal DAG, the
comparison of multiple DAGs with similar scores is a common practice, focusing
on
consistent topological features. The second step for computing the posterior
probabilities involves parameter estimation. It depends on the form of the
conditional probabilities (whether these are discrete, continuous, or involve
mixture distributions) and the presence of missing data for any node. Various
algorithms like the sum-product, Maximum Likelihood Estimation (MLE), Maximum A
Posteriori (MAP), and Expectation-Maximization (EM) are commonly applied to
address this problem \cite{friedman2000using}. Additionally, prior information
about parameter and graph distributions is incorporated into the scoring
function calculation. It is crucial for the chosen scoring function to be
decomposable into local scores for computational efficiency and to include
features that guard against overfitting. Popular strategies to achieve this
goal include the Bayesian Information Criterion (BIC) and the Bayesian
Dirichlet equivalent (BDe) \cite{yoo2001discovery,cooper1992bayesian}.

\subsection{Interologous networks}\label{subsec:iin}
The interologous network approach is a valuable strategy for constructing
networks of contigs obtained from expression data, leveraging existing
protein interaction information \cite{singh2019construction}. This method
relies on the concept of
orthology, which identifies orthologous proteins in a template organism and
subsequently transfers interactions between these orthologous proteins to
the query proteome, which consists of the proteins encoded by the contigs
obtained after assembly. The procedure begins with the selection of one or more
template organisms, typically those for which extensive protein interaction
data is available. These template organisms serve as a reference, providing a
foundation for inferring interactions in the target organism, where such data
may be limited or unavailable \cite{matthews2001identification}.

\subsubsection{Ortholog Identification}\label{subsubsec:ortho}
The first step involves identifying orthologous proteins between the template
organism(s) and the target organism. Orthologs are genes that are related by
vertical descent from a common ancestor and encode proteins with the same
function in different species \cite{fitch1970distinguishing}. Orthologous
relationships can be classified into three different types, namely, one-to-one
orthologs, one-to-many orthologs and many-to-many orthologs based on the
evolutionary events that led to their divergence. Genes that have a single copy
in each species, and are derived from a single gene in the last common ancestor
are called one-to-one orthologs. Genes having a single copy in one species, but
multiple copies in another species, due to gene duplication events after
speciation are called one-to-many orthologs. Finally genes that have multiple
copies in both species, due to gene duplication events before and after
speciation many-to-many orthologs.

In the quest to identify orthologous genes, several distinct approaches have
emerged, each with its own set of tools and techniques. These methods can be
broadly categorised into three main groups, namely, graph-based methods,
phylogenetic tree-based methods and meta or hybrid methods
\cite{kuzniar2008quest,kristensen2011computational,tekaia2016inferring}.
Graph-based approaches for orthology detection have evolved as a response to the
growing availability of complete genome sequences and the need for efficient
methods to discern orthologous genes. These methods construct graphs depicting
genes or proteins as nodes, and edges signify their evolutionary relationships.
Typically, these methods follow a two-phase process: the graph construction and
clustering phases. The graph construction phase in orthology detection is based
on the premise that orthologous genes are the homologous genes that have emerged
due to speciation events in their most recent common ancestor. This implies they
have undergone minimal divergence since they branched at the latest conceivable
juncture in their evolutionary history. Grap-based approaches employ sequence
similarity scores as a proxy for evolutionary divergence to determine
orthologous genes by identifying the highest-scoring match or genome-wide best
hit for each gene in the query genome \cite{tatusov1997genomic}. Since
the orthology relation is symmetric, reciprocal best hits, also known as
bi-directional best hits (BBH), are mostly considered \cite{overbeek1999use}.
Maximum likelihood estimates computed based on amino acid substitutions are
another commonly used measure of evolutionary distance between sequence pairs
\cite{wall2003detecting}.

Following speciation events in the two genomes under consideration, there are
instances where a gene might undergo duplication, giving rise to multiple
orthologous genes or in-paralogs \cite{remm2001automatic}. The nature of the
relationship, whether it’s one-to-many or many-to-many, hinges on whether gene
duplication occurred in one or both genomes. However, the conventional
Bi-directional Best Hit (BBH) approach is limited to predicting only one-to-one
relationships. To address this limitation, InParanoid
\cite{remm2001automatic,o2005inparanoid} identifies in-paralogous genes in the
query genome that exhibit more significant similarity to each other than the BBH
gene in the subject genome. Once pairs of orthologs have been identified, the
next step involves clustering these genes into orthologous groups. Various
clustering methods are employed for this purpose. Clusters of Orthologous Groups
(COGs) method identifies triangles in the orthologous network and proceeds by
merging triangles that share a common face, iteratively combining them until no
more triangles can be added to the existing COGs \cite{tatusov1997genomic}.
OrthoMCL employs a Markov clustering technique by simulating random walks on the
orthology graph and returns probabilities to pairs of genes belonging to the
same cluster. Subsequently, the graph is partitioned into groups based on these
probabilities \cite{li2003orthomcl}. OrthoDB \cite{waterhouse2011orthodb} and
EggNOG \cite{jensen2007eggnog} employ a combination of hierarchical clustering
and COGs-like clustering methods. In contrast, Hieranoid
\cite{schreiber2013hieranoid}, OMAGETHOG’s \cite{train2017orthologous}, and
COCO-CL \cite{jothi2006coco} generate hierarchical orthologous groups by
utilising a guide tree, taxonomic information, and correlations of similarity
scores among homologous genes, respectively.

Tree-based methods for orthology inference primarily rely on reconstructing
evolutionary history by combining the phylogenies of genes and species. A gene
tree is initially constructed using multiple sequence alignments from candidate
gene family sequences. The species tree and gene tree are then reconciled, and
the internal nodes of the gene tree are annotated with various evolutionary
processes, including speciation, loss, and duplication
\cite{tekaia2016inferring}. Predicting orthologous and paralogous relationships
is simple once these events have been identified. A parsimony criterion is used
by most tree reconciliation techniques, which favours reconciliations with the
fewest gene duplications and deletions \cite{goodman1979fitting}. Many methods
like RIO \cite{zmasek2002rio}, Orthostrapper \cite{storm2002automated},
PhylomeDB \cite{huerta2014phylomedb} implement tree reconciliation to generate
accurate hierarchical ortholog groupings. Flat lists of orthologs lacking
intra-group relationships are less informative than these hierarchical groups.
Some significant disadvantages associated with phylogenetic tree-based methods,
preventing their application to large-scale datasets, include computational
complexity and dependence on accurate multiple alignments and trees
\cite{tekaia2016inferring}.

Meta-methods amalgamate multiple orthology prediction approaches to generate a
robust consensus output. They accomplish this by intersecting the results from
several methods and assigning scores to each prediction based on the number of
independent predictors that support that relationship. Consequently, predictions
garner higher scores when they are corroborated by multiple predictors. Notably,
methods like DIPOT \cite{hu2011integrative}, GET\_HOMOLOGUES
\cite{contreras2013get_homologues}, COMPARE \cite{salgado2008compare}, and HCOP
\cite{eyre2007hcop} adopt this approach to assign weights to their predictions.
Certain methods incorporate postprocessing strategies to enhance their results.
For instance, MOSAIC \cite{maher2015rock} conducts iterative graph-based
optimization to incorporate missing orthologs effectively. MARIO
\cite{pereira2014meta} identifies common predictions from various orthology
algorithms as seeds and incrementally adds additional sequences into different
clusters using HMM profiles. OMA Hierarchical Orthologous Groups (HOG) employs
an orthology graph to form groups progressively, beginning from specific
taxonomic levels and merging towards the root of the species tree
\cite{altenhoff2013inferring}. OrthoFinder2 \cite{emms2019orthofinder} combines
graph-based and tree-based approaches, first identifying orthogroups using the
OrthoFinder graph-based method and then leveraging these to infer approximate
gene trees and species trees, enhancing orthology and paralogy predictions.
Methods such as WORMHOLE \cite{sutphin2016wormhole} employ machine learning
algorithms to recognize patterns among multiple orthology prediction methods and
subsequently leverage these patterns to identify novel orthologs.

\subsubsection{Interaction Transfer and Network
Construction}\label{subsubsec:itnc}
Once orthologs are identified, the known protein interactions from the template
organism are transferred to the target organism. This transfer assumes that if
two proteins in the template organism interact, their orthologous counterparts
in the target organism are also likely to interact
\cite{matthews2001identification,singh2019tulsipin}. Thus, transcript contigs in
the target organism are connected based on the interactions between their
corresponding orthologs in the template species \cite{singh2019construction}.
This network represents potential functional associations and regulatory
relationships among the transcript contigs. Several repositories
collect, store, and annotate protein interaction data from various sources,
such as experimental methods, computational methods, literature mining, and
manual curation. Some of the commonly used databases of PPIs are BioGRID
\cite{oughtred2019biogrid}, IntAct \cite{kerrien2012intact}, STRING
\cite{szklarczyk2021string} and MINT \cite{licata2012mint}. These databases can
provide easy access and integration of protein interaction data, but they may
have different standards, formats, and coverage. The interaction information for
the template species is obtained from these publically available repositories.

\section{Network analysis}\label{sec:na}
Building upon the foundational understanding of network construction from
transcriptomic data, we now embark on the network analysis phase. Here, we will
unravel the intricate relationships embedded within coexpression networks,
ultimately shedding light on the functional dynamics of genes and proteins
(Fig \ref{fig:MO}). It aids in the interpretation of high-dimensional
transcriptomic data, and we will explore how these networks can guide us in
extracting meaningful biological insights, and advancing our knowledge across a
spectrum of scientific domains, from fundamental research to the development of
cutting-edge medical and biotechnological innovations.

\subsection{Evaluation of interaction significance}\label{subsec:eis}
In coexpression or protein interaction networks, each interaction is associated
with a specific score, which is determined by the predictive method. This
score gives and indication of the strength of the predicted interaction,
however, this is approximate. The most authentic way to determine the
significance of an interaction is through experimental validation. However,
this process is often time-consuming, costly, and labor-intensive. As an
alternative, computational methods are available to serve this purpose.
Therefore, it is important to assess the significance of the predicted
interactions using various methods, such as topological, and biological
function based statistical methods \cite{peng2017protein}.

\subsubsection{Topology based statistical significance}\label{subsubsec:tbss}

Random networks are generated using null models such as Erdős–Rényi (ER)
\cite{erdHos1960evolution}, scale-free (SF) \cite{albert2002statistical},
Configurational \cite{newman2018networks} \emph{etc.}, and are subsequently
compared to the original network using selected topological properties. This
allows for the computation of statistical measures like \emph{z-scores},
\emph{p-values}, or false discovery rates (FDR) to assess the significance of
the predictions \cite{singh2019tulsipin}. Another approach involves random
sampling of a portion of the original network in each iteration, where the
frequencies of each edge are recorded. The edges that are observed in a greater
number of iterations are considered to be stronger or more significant
\cite{friedman2000using}. Alternatively, mutual rank (MR) can be
calculated, where various similarity or distance methods are employed to
determine MR values for each prediction \cite{obayashi2022atted}. These
statistical methods play a crucial role in evaluating the significance and
confidence of predictions, but their effectiveness relies on the validity and
robustness of the underlying models or assumptions. Nodes with associated
\emph{p-values} less than a predefined threshold, typically 0.01, in both the ER
and SF models are considered key nodes or drivers.

\subsubsection{Biological function based statistical
significance}\label{subsubsec:bfbs}

Another approach to assess significance is by leveraging existing biological
knowledge related to pathways, cellular localization, and gene ontology. It's a
common belief that interacting proteins often participate in the same pathways
and share cellular locations. To evaluate this, information about pathways and
cellular locations for the interacting proteins is gathered, and their
congruence is analyzed. Gene ontology provides valuable insights into cellular
processes, molecular functions, and subcellular localization of proteins, making
it a useful resource for assessing interaction significance
\cite{singh2019tulsipin}. Typically, quantitative scores for predictions are
obtained using metrics like the Jaccard similarity index or various semantic
similarity indices to measure the similarity \cite{yu2010gosemsim} between the
annotated attributes of the proteins.

\subsection{Functional modules identification}\label{subsec:fmi}

Coexpression data provides the expression levels of genes or proteins
under different conditions, and gives insights into additional
information to infer
the functional modules, as proteins that are coexpressed are likely to be
functionally related. Categorising various module detection methods can be
challenging due to the fine boundaries between them; for instance, matrix
decomposition is an intermediate step in some clustering and biclustering
algorithms. These methods are generally grouped into three categories:
clustering, decomposition and biclustering methods
\cite{saelens2018comprehensive}. Clustering methods, among the oldest and still
widely used, group genes based on their overall similarity in gene expression.
Commonly applied clustering algorithms for module detection include K-medoids
\cite{schubert2021fast}, K-means \cite{hartigan1979algorithm}, Fuzzy-c-means
\cite{bezdek1984fcm}, MCL \cite{van2000graph}, Agglomerative hierarchical
clustering \cite{kaufman2009finding}, and WGCNA \cite{langfelder2008wgcna}.
Decomposition methods aim to approximate the expression matrix through the
product of smaller matrices. Within this approximation, two matrices hold the
individual contributions of genes and samples to specific modules. By limiting
the extent to which samples can contribute to a module, decomposition methods
are effective at detecting local coexpression patterns. Commonly employed
decomposition-based methods encompass Independent Component Analysis (ICA)
\cite{hyvarinen1999fast}, Principal Component Analysis (PCA)
\cite{pedregosa2011scikit}, and hybrid approaches that merge elements of both
techniques \cite{yao2012independent}. Biclustering methods excel at identifying
clusters of genes and samples that exhibit localised coexpression exclusively
within the confines of the bicluster. Unlike decomposition methods, where all
samples contribute to some extent, biclustering categorises samples as either
contributing or not to a specific module. Consequently, modules unveiled by
biclustering methods are often more interpretable as they provide a more
precise delineation of the exact source of local coexpression. In some
instances, a biclustering approach can be considered an extension of an existing
decomposition method, but with the added requirement that gene and sample
contributions to a module are sparse, containing numerous zeros. Spectral
biclustering \cite{kluger2003spectral}, Iterative Signature Algorithm (ISA)
\cite{ihmels2002revealing}, QUalitative BIClustering algorithm (QUBIC)
\cite{li2009qubic}, Bi-Force \cite{sun2014bi}, and Factor Analysis for BIcluster
Acquisition (FABIA) \cite{hochreiter2010fabia} are some of the commonly
utilised biclustering techniques.

\subsection{Identification of key drivers}\label{subsec:ikd}

Identifying key drivers within coexpression or protein interaction networks is a
fundamental task in systems biology \cite{carter2004gene}. These identified key
drivers within the network often correspond to proteins with critical roles in
various pathways \cite{wisecaver2017global}, or they may code for transcription
factors that regulate gene expression \cite{haynes2013mapping}. Thus are
crucial for understanding the underlying regulatory mechanisms and functional
roles of genes or proteins. Several methodologies are employed for this
purpose, ranging from straightforward topological metrics to more sophisticated
statistical analyses.

\subsubsection{Centrality based methods}\label{subsubsec:cbm}
One of the simplest approaches involves using topological metrics to assess the
importance of nodes within the network \cite{koschutzki2008centrality}. These
metrics evaluate characteristics such as degree centrality
\cite{jeong2001lethality}, betweenness centrality \cite{joy2005high}, and
closeness centrality \cite{wuchty2002interaction}, which respectively measure a
node’s connectedness, its position in facilitating information flow, and its
proximity to other nodes. Nodes with high values of these metrics are
considered key drivers, as they play pivotal roles in network connectivity and
communication \cite{koschutzki2008centrality}. There exists a diverse array of
centrality measures, numbering over 200 \cite{jalili2015centiserver}, each
meticulously crafted to capture distinct facets of network topology.
Consequently, a fusion of various centrality measures has been employed to
unveil nodes of significance within networks. This amalgamation enables a more
comprehensive understanding of node importance by considering a spectrum of
network characteristics
\cite{ma2003connectivity,lareau2015differential}. For instance, by generating
numerous realisations of these random networks, researchers compute topological
metrics for the original network and the random ensembles. Statistical
significance scores, often expressed as z-scores, are then calculated for each
node within the original network. Nodes with associated \emph{p-values} less
than a
predefined threshold, typically 0.01, are considered key nodes or drivers
\cite{singh2019tulsipin}.

\subsubsection{Diffusion based methods}\label{subsubsec:dbm}

Diffusion based methods for node prioritization are methods that simulate the
spread of information or influence along paths within the network. They
prioritize nodes based on their ability to propagate information to other nodes.
Random walk based methods stand as a prominent class of diffusion-based node
prioritization techniques, offering a means to gauge node importance in a
network \cite{ata2021recent}. These methods leverage random walks, stochastic
processes that emulate the meandering path of a random surfer navigating a
graph. The surfer initiates the journey from a source node and, at each step,
selects one of its neighbors to visit randomly. The likelihood of choosing a
particular node hinges on the network's structural properties and edge weights.
Node ranking in random walk-based methods is contingent upon the probability of
being visited by the random surfer, which effectively represents a node's
proximity or significance concerning the source node. Examples of such methods
include Random Walk with Restart (RWR) \cite{kohler2008walking}, which gauges
the likelihood of reaching each node from the source node through random walks
with a restart probability. Personalized PageRank
\cite{brin1998pagerank,wang2020personalized}, akin to RWR but customized to a
specific set of query nodes, evaluates node importance based on their closeness
to the query nodes. Diffusion Component Analysis (DCA) \cite{cho2015diffusion}
identifies influential nodes by assessing their contributions to the diffusion
process and utilizes matrix decomposition techniques to unearth significant
nodes in information propagation.

\subsubsection{Subgraph based methods}\label{subsubsec:sbm}

Subgraph based node prioritisation methods are grounded in using subgraphs as
fundamental units for network analysis and representation. These techniques
assess the importance of nodes based on their roles and memberships within these
subgraphs, which represent coherent structures in the network, such as
communities, clusters, or motifs. Subgraph based approaches possess the
flexibility to capture both local and global patterns within the network,
thereby excelling in the identification of critical nodes that facilitate
connections both between subgraphs and within them. Common subgraph
based methodologies encompass module detection
\cite{ideker2002discovering,saelens2018comprehensive}, which identifies
functionally related node groups; motif discovery
\cite{milo2002network,prvzulj2004modeling}, which detects recurring structural
patterns; community detection \cite{newman2004finding}, which pinpoints densely
connected node groups; and subgraph embedding, which maps nodes into vector
spaces considering their subgraph structures
\cite{alsentzer2020subgraph,adhikari2018sub2vec}.

\section{Multistep protocol for interaction network construction and analysis}

\begin{enumerate}
    \item \textbf{Data Preprocessing:} Begin by collecting gene expression data
from relevant experiments or databases. Clean this data and remove outliers and
noise to ensure data integrity. Following this, apply normalisation to reconcile
the data, accounting for sample variations and setting the stage for consistent
analysis (see Section \ref{sec:pot} for details).
    \item \textbf{Network Construction:} Follow one of the three abovementioned
methodologies in Section \ref{sec:iin}. to construct the network. For
association-based methods, calculate correlations or mutual information scores
between gene expressions. Apply Gaussian graphical models or the Bayesian
approach to infer the probabilistic network structure. Additionally, integrate
protein-protein interaction data through interologous method, mapping gene
expression onto the interactome to construct a comprehensive network.
    \item \textbf{Reliability assessment of predicted interactions:}
Cross-validate the constructed network against known biological interactions or
experimental data for reliability. Apply topological and biological
function-based statistical significance assessment methods mentioned in Section
\ref{subsec:eis} to ensure that predictions are significant.
    \item \textbf{Network Visualisation:} Use network visualisation tools like
Cytoscape \cite{shannon2003cytoscape} or Gephi \cite{bastian2009gephi} to
visualise and interpret the network effectively. Highlight nodes based on
expression levels, using distinct shapes or colours for key nodes or modules.
This step ensures that the intricate network structures are not only analysed
but also visually accessible and understandable.
    \item \textbf{Results Interpretation and Conclusion:} Now, synthesise and
interpret the findings from network analysis. Summarise the modules (see Section
\ref{subsec:fmi}), key nodes (see Section \ref{subsec:ikd}), and enriched
functions, and discuss their biological implications and significance. This
step provides a comprehensive understanding of the biological insights from the
constructed and analysed network.
\end{enumerate}

\section{Summary and future prospectives}

In summary, this chapter has delved into the realm of network inference from
transcriptomic data, highlighting the pivotal role of two primary datasets:
microarrays and RNA sequencing (RNA-seq). While microarrays were once the go-to
technology, RNA-seq has surged ahead due to its increased sophistication and
versatility. The construction of coexpression networks and interologous networks
of contigs was thoroughly explored, offering insights into how these networks
can reveal intricate biological relationships. Moreover, network analysis
emerged as a central theme, encompassing the assessment of interaction
significance, identifying functional modules, and discovering essential driver
genes. These analytical processes shed light on the complex interplay within
biological systems and provide a deeper understanding of gene regulation and
function. Looking ahead, the field of network inference from transcriptomic data
holds promising prospects. With the continuous advancements in high-throughput
sequencing technologies, we can anticipate even more comprehensive and precise
transcriptomic datasets. Integrative multi-omics approaches that combine
multiple data types, such as genomics and proteomics, are likely to become
increasingly prevalent, enabling a more holistic understanding of cellular
processes. Furthermore, the development of novel algorithms and computational
techniques will play a pivotal role in refining network inference and analysis.
These innovations will empower researchers to extract deeper insights from
complex biological data, unravelling the intricacies of gene regulatory networks
and contributing to breakthroughs in fields like systems biology and
personalised medicine. In conclusion, as we navigate the ever-evolving landscape
of transcriptomic data analysis, the future of network inference holds immense
potential for uncovering the mysteries of cellular dynamics and advancing our
knowledge of biological systems.

%BibTeX users: After compilation, comment out the following two lines and paste
% in the generated .bbl file.

\bibliography{scibib_mod}

\begin{thebibliography}{100}

\bibitem{crick1970central}
F.~Crick, ``{Central dogma of molecular biology},'' {\em Nature}, vol.~227,
  no.~5258, pp.~561--563, 1970.

\bibitem{bustamante2011revisiting}
C.~Bustamante, W.~Cheng, and Y.~X. Mejia, ``{Revisiting the central dogma one
  molecule at a time},'' {\em Cell}, vol.~144, no.~4, pp.~480--497, 2011.

\bibitem{jacob1961genetic}
F.~Jacob and J.~Monod, ``{Genetic regulatory mechanisms in the synthesis of
  proteins},'' {\em Journal of Molecular Biology}, vol.~3, no.~3, pp.~318--356,
  1961.

\bibitem{stark1978ribonuclease}
B.~C. Stark, R.~Kole, E.~J. Bowman, and S.~Altman, ``{Ribonuclease P: an enzyme
  with an essential RNA component.},'' {\em Proceedings of the National Academy
  of Sciences}, vol.~75, no.~8, pp.~3717--3721, 1978.

\bibitem{morris2014rise}
K.~V. Morris and J.~S. Mattick, ``{The rise of regulatory RNA},'' {\em Nature
  Reviews Genetics}, vol.~15, no.~6, pp.~423--437, 2014.

\bibitem{siva20081000}
N.~Siva, ``{1000 Genomes project},'' {\em Nature Biotechnology}, vol.~26,
  no.~3, pp.~256--257, 2008.

\bibitem{segal2002decomposing}
E.~Segal, A.~Battle, and D.~Koller, ``{Decomposing gene expression into
  cellular processes},'' in {\em Biocomputing 2003}, pp.~89--100, World
  Scientific, 2002.

\bibitem{lowe2017transcriptomics}
R.~Lowe, N.~Shirley, M.~Bleackley, S.~Dolan, and T.~Shafee, ``{Transcriptomics
  technologies},'' {\em Plos Computational Biology}, vol.~13, no.~5,
  p.~e1005457, 2017.

\bibitem{emilsson2008genetics}
V.~Emilsson, G.~Thorleifsson, B.~Zhang, A.~S. Leonardson, F.~Zink, J.~Zhu,
  S.~Carlson, A.~Helgason, G.~B. Walters, S.~Gunnarsdottir, {\em et~al.},
  ``{Genetics of gene expression and its effect on disease},'' {\em Nature},
  vol.~452, no.~7186, pp.~423--428, 2008.

\bibitem{maloy1993autogenous}
S.~Maloy and V.~Stewart, ``{Autogenous regulation of gene expression},'' {\em
  Journal of Bacteriology}, vol.~175, no.~2, pp.~307--316, 1993.

\bibitem{maniatis1987regulation}
T.~Maniatis, S.~Goodbourn, and J.~A. Fischer, ``{Regulation of inducible and
  tissue-specific gene expression},'' {\em Science}, vol.~236, no.~4806,
  pp.~1237--1245, 1987.

\bibitem{killary1984genetic}
A.~Killary and R.~Fournier, ``{A genetic analysis of extinction: trans-dominant
  loci regulate expression of liver-specific traits in hepatoma hybrid
  cells},'' {\em Cell}, vol.~38, no.~2, pp.~523--534, 1984.

\bibitem{wen1998large}
X.~Wen, S.~Fuhrman, G.~S. Michaels, D.~B. Carr, S.~Smith, J.~L. Barker, and
  R.~Somogyi, ``{Large-scale temporal gene expression mapping of central
  nervous system development},'' {\em Proceedings of the National Academy of
  Sciences}, vol.~95, no.~1, pp.~334--339, 1998.

\bibitem{carninci2005transcriptional}
P.~Carninci, T.~Kasukawa, S.~Katayama, J.~Gough, M.~Frith, N.~Maeda, R.~Oyama,
  T.~Ravasi, B.~Lenhard, C.~Wells, {\em et~al.}, ``{The transcriptional
  landscape of the mammalian genome},'' {\em Science}, vol.~309, no.~5740,
  pp.~1559--1563, 2005.

\bibitem{jacquier2009complex}
A.~Jacquier, ``{The complex eukaryotic transcriptome: unexpected pervasive
  transcription and novel small RNAs},'' {\em Nature Reviews Genetics},
  vol.~10, no.~12, pp.~833--844, 2009.

\bibitem{marra1998expressed}
M.~A. Marra, L.~Hillier, and R.~H. Waterston, ``{Expressed sequence
  tags—ESTablishing bridges between genomes},'' {\em Trends in Genetics},
  vol.~14, no.~1, pp.~4--7, 1998.

\bibitem{sanger1977dna}
F.~Sanger, S.~Nicklen, and A.~R. Coulson, ``{DNA sequencing with
  chain-terminating inhibitors},'' {\em Proceedings of the National Academy of
  Sciences}, vol.~74, no.~12, pp.~5463--5467, 1977.

\bibitem{velculescu1995serial}
V.~E. Velculescu, L.~Zhang, B.~Vogelstein, and K.~W. Kinzler, ``{Serial
  analysis of gene expression},'' {\em Science}, vol.~270, no.~5235,
  pp.~484--487, 1995.

\bibitem{schena1995quantitative}
M.~Schena, D.~Shalon, R.~W. Davis, and P.~O. Brown, ``{Quantitative monitoring
  of gene expression patterns with a complementary DNA microarray},'' {\em
  Science}, vol.~270, no.~5235, pp.~467--470, 1995.

\bibitem{ozsolak2011rna}
F.~Ozsolak and P.~M. Milos, ``{RNA sequencing: advances, challenges and
  opportunities},'' {\em Nature Reviews Genetics}, vol.~12, no.~2, pp.~87--98,
  2011.

\bibitem{shalon1996dna}
D.~Shalon, S.~J. Smith, and P.~O. Brown, ``{A DNA microarray system for
  analyzing complex DNA samples using two-color fluorescent probe
  hybridization.},'' {\em Genome Research}, vol.~6, no.~7, pp.~639--645, 1996.

\bibitem{lockhart1996expression}
D.~J. Lockhart, H.~Dong, M.~C. Byrne, M.~T. Follettie, M.~V. Gallo, M.~S. Chee,
  M.~Mittmann, C.~Wang, M.~Kobayashi, H.~Norton, {\em et~al.}, ``{Expression
  monitoring by hybridization to high-density oligonucleotide arrays},'' {\em
  Nature Biotechnology}, vol.~14, no.~13, pp.~1675--1680, 1996.

\bibitem{wang2009rna}
Z.~Wang, M.~Gerstein, and M.~Snyder, ``{RNA-Seq: a revolutionary tool for
  transcriptomics},'' {\em Nature Reviews Genetics}, vol.~10, no.~1,
  pp.~57--63, 2009.

\bibitem{van2013rna}
M.~C. Van~Verk, R.~Hickman, C.~M. Pieterse, and S.~C. Van~Wees, ``{RNA-Seq:
  revelation of the messengers},'' {\em Trends in Plant Science}, vol.~18,
  no.~4, pp.~175--179, 2013.

\bibitem{huber2015orchestrating}
W.~Huber, V.~J. Carey, R.~Gentleman, S.~Anders, M.~Carlson, B.~S. Carvalho,
  H.~C. Bravo, S.~Davis, L.~Gatto, T.~Girke, {\em et~al.}, ``{Orchestrating
  high-throughput genomic analysis with Bioconductor},'' {\em Nature Methods},
  vol.~12, no.~2, pp.~115--121, 2015.

\bibitem{amezquita2020orchestrating}
R.~A. Amezquita, A.~T. Lun, E.~Becht, V.~J. Carey, L.~N. Carpp, L.~Geistlinger,
  F.~Marini, K.~Rue-Albrecht, D.~Risso, C.~Soneson, {\em et~al.},
  ``{Orchestrating single-cell analysis with Bioconductor},'' {\em Nature
  Methods}, vol.~17, no.~2, pp.~137--145, 2020.

\bibitem{macmahon1978levels}
J.~A. MacMahon, D.~L. Phillips, J.~V. Robinson, and D.~J. Schimpf, ``{Levels of
  biological organization: an organism-centered approach},'' {\em Bioscience},
  vol.~28, no.~11, pp.~700--704, 1978.

\bibitem{newman2003structure}
M.~E. Newman, ``{The structure and function of complex networks},'' {\em Siam
  Review}, vol.~45, no.~2, pp.~167--256, 2003.

\bibitem{albert2002statistical}
R.~Albert and A.-L. Barab{\'a}si, ``{Statistical mechanics of complex
  networks},'' {\em Reviews of Modern Physics}, vol.~74, no.~1, p.~47, 2002.

\bibitem{bascompte2009disentangling}
J.~Bascompte, ``{Disentangling the web of life},'' {\em Science}, vol.~325,
  no.~5939, pp.~416--419, 2009.

\bibitem{gardner2003inferring}
T.~S. Gardner, D.~Di~Bernardo, D.~Lorenz, and J.~J. Collins, ``{Inferring
  genetic networks and identifying compound mode of action via expression
  profiling},'' {\em Science}, vol.~301, no.~5629, pp.~102--105, 2003.

\bibitem{d2000genetic}
P.~D’haeseleer, S.~Liang, and R.~Somogyi, ``{Genetic network inference: from
  co-expression clustering to reverse engineering},'' {\em Bioinformatics},
  vol.~16, no.~8, pp.~707--726, 2000.

\bibitem{wang2014review}
Y.~R. Wang and H.~Huang, ``{Review on statistical methods for gene network
  reconstruction using expression data},'' {\em Journal of Theoretical
  Biology}, vol.~362, pp.~53--61, 2014.

\bibitem{friedman2000using}
N.~Friedman, M.~Linial, I.~Nachman, and D.~Pe'er, ``{Using Bayesian networks to
  analyze expression data},'' in {\em Proceedings of the fourth annual
  international conference on Computational molecular biology}, pp.~127--135,
  2000.

\bibitem{horvath2008geometric}
S.~Horvath and J.~Dong, ``{Geometric interpretation of gene coexpression
  network analysis},'' {\em Plos Computational Biology}, vol.~4, no.~8,
  p.~e1000117, 2008.

\bibitem{estrada2012structure}
E.~Estrada, {\em {The structure of complex networks: theory and applications}}.
\newblock Oxford University Press, USA, 2012.

\bibitem{assenov2008computing}
Y.~Assenov, F.~Ram{\'\i}rez, S.-E. Schelhorn, T.~Lengauer, and M.~Albrecht,
  ``{Computing topological parameters of biological networks},'' {\em
  Bioinformatics}, vol.~24, no.~2, pp.~282--284, 2008.

\bibitem{wang2010degseq}
L.~Wang, Z.~Feng, X.~Wang, X.~Wang, and X.~Zhang, ``{DEGseq: an R package for
  identifying differentially expressed genes from RNA-seq data},'' {\em
  Bioinformatics}, vol.~26, no.~1, pp.~136--138, 2010.

\bibitem{kvam2012comparison}
V.~M. Kvam, P.~Liu, and Y.~Si, ``{A comparison of statistical methods for
  detecting differentially expressed genes from RNA-seq data},'' {\em American
  Journal of Botany}, vol.~99, no.~2, pp.~248--256, 2012.

\bibitem{slonim2002patterns}
D.~K. Slonim, ``{From patterns to pathways: gene expression data analysis comes
  of age},'' {\em Nature Genetics}, vol.~32, no.~4, pp.~502--508, 2002.

\bibitem{jain1988algorithms}
A.~K. Jain and R.~C. Dubes, {\em {Algorithms for clustering data}}.
\newblock Prentice-Hall, Inc., 1988.

\bibitem{raychaudhuri2000principal}
S.~Raychaudhuri, J.~M. Stuart, and R.~B. Altman, ``{Principal components
  analysis to summarize microarray experiments: application to sporulation time
  series},'' in {\em Biocomputing 2000}, pp.~455--466, World Scientific, 2000.

\bibitem{pellegrini2004protein}
M.~Pellegrini, D.~Haynor, and J.~M. Johnson, ``{Protein interaction
  networks},'' {\em Expert Review of Proteomics}, vol.~1, no.~2, pp.~239--249,
  2004.

\bibitem{kanehisa2000kegg}
M.~Kanehisa and S.~Goto, ``{KEGG: kyoto encyclopedia of genes and genomes},''
  {\em Nucleic Acids Research}, vol.~28, no.~1, pp.~27--30, 2000.

\bibitem{van2018gene}
S.~Van~Dam, U.~Vosa, A.~van~der Graaf, L.~Franke, and J.~P. de~Magalhaes,
  ``{Gene co-expression analysis for functional classification and
  gene--disease predictions},'' {\em Briefings in Bioinformatics}, vol.~19,
  no.~4, pp.~575--592, 2018.

\bibitem{langfelder2008wgcna}
P.~Langfelder and S.~Horvath, ``{WGCNA: an R package for weighted correlation
  network analysis},'' {\em Bmc Bioinformatics}, vol.~9, no.~1, pp.~1--13,
  2008.

\bibitem{pearson1895vii}
K.~Pearson, ``{VII. Note on regression and inheritance in the case of two
  parents},'' {\em Proceedings of the Royal Society of London}, vol.~58,
  no.~347-352, pp.~240--242, 1895.

\bibitem{kendall1938new}
M.~G. Kendall, ``{A new measure of rank correlation},'' {\em Biometrika},
  vol.~30, no.~1/2, pp.~81--93, 1938.

\bibitem{spearman1961proof}
C.~Spearman, ``{The proof and measurement of association between two
  things.},'' 1961.

\bibitem{song2012comparison}
L.~Song, P.~Langfelder, and S.~Horvath, ``{Comparison of co-expression
  measures: mutual information, correlation, and model based indices},'' {\em
  Bmc Bioinformatics}, vol.~13, no.~1, pp.~1--21, 2012.

\bibitem{kumari2012evaluation}
S.~Kumari, J.~Nie, H.-S. Chen, H.~Ma, R.~Stewart, X.~Li, M.-Z. Lu, W.~M.
  Taylor, and H.~Wei, ``{Evaluation of gene association methods for
  coexpression network construction and biological knowledge discovery},'' {\em
  Plos One}, vol.~7, no.~11, p.~e50411, 2012.

\bibitem{butte1999mutual}
A.~J. Butte and I.~S. Kohane, ``{Mutual information relevance networks:
  functional genomic clustering using pairwise entropy measurements},'' in {\em
  Biocomputing 2000}, pp.~418--429, World Scientific, 1999.

\bibitem{lopez2022network}
N.~L{\'o}pez-Rozo, M.~Romero, J.~Finke, and C.~Rocha, ``{A Network-based
  Approach for Inferring Thresholds in Co-expression Networks},'' in {\em
  International Conference on Complex Networks and Their Applications},
  pp.~265--276, Springer, 2022.

\bibitem{paci2017swim}
P.~Paci, T.~Colombo, G.~Fiscon, A.~Gurtner, G.~Pavesi, and L.~Farina, ``{SWIM:
  a computational tool to unveiling crucial nodes in complex biological
  networks},'' {\em Scientific Reports}, vol.~7, no.~1, p.~44797, 2017.

\bibitem{butte2000determining}
A.~J. Butte, J.~Ye, H.~H{\"a}ring, M.~Stumvoll, M.~White, and I.~Kohane,
  ``{Determining significant fold differences in gene expression analysis},''
  in {\em Biocomputing 2001}, pp.~6--17, World Scientific, 2000.

\bibitem{carter2004gene}
S.~L. Carter, C.~M. Brechb{\"u}hler, M.~Griffin, and A.~T. Bond, ``{Gene
  co-expression network topology provides a framework for molecular
  characterization of cellular state},'' {\em Bioinformatics}, vol.~20, no.~14,
  pp.~2242--2250, 2004.

\bibitem{zhang2005general}
B.~Zhang and S.~Horvath, ``{A general framework for weighted gene co-expression
  network analysis},'' {\em Statistical Applications in Genetics and Molecular
  Biology}, vol.~4, no.~1, 2005.

\bibitem{schafer2005empirical}
J.~Sch{\"a}fer and K.~Strimmer, ``{An empirical Bayes approach to inferring
  large-scale gene association networks},'' {\em Bioinformatics}, vol.~21,
  no.~6, pp.~754--764, 2005.

\bibitem{edwards2012introduction}
D.~Edwards, {\em {Introduction to graphical modelling}}.
\newblock Springer Science \& Business Media, 2012.

\bibitem{friedman2008sparse}
J.~Friedman, T.~Hastie, and R.~Tibshirani, ``{Sparse inverse covariance
  estimation with the graphical lasso},'' {\em Biostatistics}, vol.~9, no.~3,
  pp.~432--441, 2008.

\bibitem{pearl1988probabilistic}
J.~Pearl, {\em {Probabilistic reasoning in intelligent systems: networks of
  plausible inference}}.
\newblock Morgan kaufmann, 1988.

\bibitem{needham2007primer}
C.~J. Needham, J.~R. Bradford, A.~J. Bulpitt, and D.~R. Westhead, ``{A primer
  on learning in Bayesian networks for computational biology},'' {\em Plos
  Computational Biology}, vol.~3, no.~8, p.~e129, 2007.

\bibitem{yu2002using}
J.~Yu, V.~A. Smith, P.~P. Wang, A.~J. Hartemink, and E.~D. Jarvis, ``{Using
  Bayesian network inference algorithms to recover molecular genetic regulatory
  networks},'' in {\em International Conference on Systems Biology}, vol.~2002,
  2002.

\bibitem{yoo2001discovery}
C.~Yoo, V.~Thorsson, and G.~F. Cooper, ``{Discovery of causal relationships in
  a gene-regulation pathway from a mixture of experimental and observational
  DNA microarray data},'' in {\em Biocomputing 2002}, pp.~498--509, World
  Scientific, 2001.

\bibitem{cooper1992bayesian}
G.~F. Cooper and E.~Herskovits, ``{A Bayesian method for the induction of
  probabilistic networks from data},'' {\em Machine Learning}, vol.~9,
  pp.~309--347, 1992.

\bibitem{singh2019construction}
G.~Singh, V.~Singh, and V.~Singh, ``{Construction and analysis of an
  interologous protein--protein interaction network of Camellia sinensis leaf
  (TeaLIPIN) from RNA--Seq data sets},'' {\em Plant Cell Reports}, vol.~38,
  pp.~1249--1262, 2019.

\bibitem{matthews2001identification}
L.~R. Matthews, P.~Vaglio, J.~Reboul, H.~Ge, B.~P. Davis, J.~Garrels,
  S.~Vincent, and M.~Vidal, ``{Identification of potential interaction networks
  using sequence-based searches for conserved protein-protein interactions or
  “interologs”},'' {\em Genome Research}, vol.~11, no.~12, pp.~2120--2126,
  2001.

\bibitem{fitch1970distinguishing}
W.~M. Fitch, ``{Distinguishing homologous from analogous proteins},'' {\em
  Systematic Zoology}, vol.~19, no.~2, pp.~99--113, 1970.

\bibitem{kuzniar2008quest}
A.~Kuzniar, R.~C. van Ham, S.~Pongor, and J.~A. Leunissen, ``{The quest for
  orthologs: finding the corresponding gene across genomes},'' {\em Trends in
  Genetics}, vol.~24, no.~11, pp.~539--551, 2008.

\bibitem{kristensen2011computational}
D.~M. Kristensen, Y.~I. Wolf, A.~R. Mushegian, and E.~V. Koonin,
  ``{Computational methods for Gene Orthology inference},'' {\em Briefings in
  Bioinformatics}, vol.~12, no.~5, pp.~379--391, 2011.

\bibitem{tekaia2016inferring}
F.~Tekaia, ``{Inferring orthologs: open questions and perspectives},'' {\em
  Genomics Insights}, vol.~9, pp.~GEI--S37925, 2016.

\bibitem{tatusov1997genomic}
R.~L. Tatusov, E.~V. Koonin, and D.~J. Lipman, ``{A genomic perspective on
  protein families},'' {\em Science}, vol.~278, no.~5338, pp.~631--637, 1997.

\bibitem{overbeek1999use}
R.~Overbeek, M.~Fonstein, M.~D’souza, G.~D. Pusch, and N.~Maltsev, ``{The use
  of gene clusters to infer functional coupling},'' {\em Proceedings of the
  National Academy of Sciences}, vol.~96, no.~6, pp.~2896--2901, 1999.

\bibitem{wall2003detecting}
D.~Wall, H.~Fraser, and A.~Hirsh, ``{Detecting putative orthologs},'' {\em
  Bioinformatics}, vol.~19, no.~13, pp.~1710--1711, 2003.

\bibitem{remm2001automatic}
M.~Remm, C.~E. Storm, and E.~L. Sonnhammer, ``{Automatic clustering of
  orthologs and in-paralogs from pairwise species comparisons},'' {\em Journal
  of Molecular Biology}, vol.~314, no.~5, pp.~1041--1052, 2001.

\bibitem{o2005inparanoid}
K.~P. O'brien, M.~Remm, and E.~L. Sonnhammer, ``{Inparanoid: a comprehensive
  database of eukaryotic orthologs},'' {\em Nucleic Acids Research}, vol.~33,
  no.~suppl\_1, pp.~D476--D480, 2005.

\bibitem{li2003orthomcl}
L.~Li, C.~J. Stoeckert, and D.~S. Roos, ``{OrthoMCL: identification of ortholog
  groups for eukaryotic genomes},'' {\em Genome Research}, vol.~13, no.~9,
  pp.~2178--2189, 2003.

\bibitem{waterhouse2011orthodb}
R.~M. Waterhouse, E.~M. Zdobnov, F.~Tegenfeldt, J.~Li, and E.~V. Kriventseva,
  ``{OrthoDB: the hierarchical catalog of eukaryotic orthologs in 2011},'' {\em
  Nucleic Acids Research}, vol.~39, no.~suppl\_1, pp.~D283--D288, 2011.

\bibitem{jensen2007eggnog}
L.~J. Jensen, P.~Julien, M.~Kuhn, C.~von Mering, J.~Muller, T.~Doerks, and
  P.~Bork, ``{eggNOG: automated construction and annotation of orthologous
  groups of genes},'' {\em Nucleic Acids Research}, vol.~36, no.~suppl\_1,
  pp.~D250--D254, 2007.

\bibitem{schreiber2013hieranoid}
F.~Schreiber and E.~L. Sonnhammer, ``{Hieranoid: hierarchical orthology
  inference},'' {\em Journal of Molecular Biology}, vol.~425, no.~11,
  pp.~2072--2081, 2013.

\bibitem{train2017orthologous}
C.-M. Train, N.~M. Glover, G.~H. Gonnet, A.~M. Altenhoff, and C.~Dessimoz,
  ``{Orthologous Matrix (OMA) algorithm 2.0: more robust to asymmetric
  evolutionary rates and more scalable hierarchical orthologous group
  inference},'' {\em Bioinformatics}, vol.~33, no.~14, pp.~i75--i82, 2017.

\bibitem{jothi2006coco}
R.~Jothi, E.~Zotenko, A.~Tasneem, and T.~M. Przytycka, ``{COCO-CL: hierarchical
  clustering of homology relations based on evolutionary correlations},'' {\em
  Bioinformatics}, vol.~22, no.~7, pp.~779--788, 2006.

\bibitem{goodman1979fitting}
M.~Goodman, J.~Czelusniak, G.~W. Moore, A.~E. Romero-Herrera, and G.~Matsuda,
  ``{Fitting the gene lineage into its species lineage, a parsimony strategy
  illustrated by cladograms constructed from globin sequences},'' {\em
  Systematic Biology}, vol.~28, no.~2, pp.~132--163, 1979.

\bibitem{zmasek2002rio}
C.~M. Zmasek and S.~R. Eddy, ``{RIO: analyzing proteomes by automated
  phylogenomics using resampled inference of orthologs},'' {\em Bmc
  Bioinformatics}, vol.~3, no.~1, pp.~1--19, 2002.

\bibitem{storm2002automated}
C.~E. Storm and E.~L. Sonnhammer, ``{Automated ortholog inference from
  phylogenetic trees and calculation of orthology reliability},'' {\em
  Bioinformatics}, vol.~18, no.~1, pp.~92--99, 2002.

\bibitem{huerta2014phylomedb}
J.~Huerta-Cepas, S.~Capella-Gutierrez, L.~P. Pryszcz, M.~Marcet-Houben, and
  T.~Gabaldon, ``{PhylomeDB v4: zooming into the plurality of evolutionary
  histories of a genome},'' {\em Nucleic Acids Research}, vol.~42, no.~D1,
  pp.~D897--D902, 2014.

\bibitem{hu2011integrative}
Y.~Hu, I.~Flockhart, A.~Vinayagam, C.~Bergwitz, B.~Berger, N.~Perrimon, and
  S.~E. Mohr, ``{An integrative approach to ortholog prediction for
  disease-focused and other functional studies},'' {\em Bmc Bioinformatics},
  vol.~12, pp.~1--16, 2011.

\bibitem{contreras2013get_homologues}
B.~Contreras-Moreira and P.~Vinuesa, ``{GET\_HOMOLOGUES, a versatile software
  package for scalable and robust microbial pangenome analysis},'' {\em Applied
  and Environmental Microbiology}, vol.~79, no.~24, pp.~7696--7701, 2013.

\bibitem{salgado2008compare}
D.~Salgado, G.~Gimenez, F.~Coulier, and C.~Marcelle, ``{COMPARE, a
  multi-organism system for cross-species data comparison and transfer of
  information},'' {\em Bioinformatics}, vol.~24, no.~3, pp.~447--449, 2008.

\bibitem{eyre2007hcop}
T.~A. Eyre, M.~W. Wright, M.~J. Lush, and E.~A. Bruford, ``{HCOP: a searchable
  database of human orthology predictions},'' {\em Briefings in
  Bioinformatics}, vol.~8, no.~1, pp.~2--5, 2007.

\bibitem{maher2015rock}
M.~C. Maher and R.~D. Hernandez, ``{Rock, paper, scissors: harnessing
  complementarity in ortholog detection methods improves comparative genomic
  inference},'' {\em G3: Genes, Genomes, Genetics}, vol.~5, no.~4,
  pp.~629--638, 2015.

\bibitem{pereira2014meta}
C.~Pereira, A.~Denise, and O.~Lespinet, ``{A meta-approach for improving the
  prediction and the functional annotation of ortholog groups},'' {\em Bmc
  Genomics}, vol.~15, no.~6, pp.~1--8, 2014.

\bibitem{altenhoff2013inferring}
A.~M. Altenhoff, M.~Gil, G.~H. Gonnet, and C.~Dessimoz, ``{Inferring
  hierarchical orthologous groups from orthologous gene pairs},'' {\em Plos
  One}, vol.~8, no.~1, p.~e53786, 2013.

\bibitem{emms2019orthofinder}
D.~M. Emms and S.~Kelly, ``{OrthoFinder: phylogenetic orthology inference for
  comparative genomics},'' {\em Genome Biology}, vol.~20, pp.~1--14, 2019.

\bibitem{sutphin2016wormhole}
G.~L. Sutphin, J.~M. Mahoney, K.~Sheppard, D.~O. Walton, and R.~Korstanje,
  ``{WORMHOLE: novel least diverged ortholog prediction through machine
  learning},'' {\em Plos Computational Biology}, vol.~12, no.~11, p.~e1005182,
  2016.

\bibitem{singh2019tulsipin}
V.~Singh, G.~Singh, and V.~Singh, ``Tulsipin: an interologous protein
  interactome of ocimum tenuiflorum,'' {\em Journal of proteome research},
  vol.~19, no.~2, pp.~884--899, 201 publisher={ACS Publications}.

\bibitem{oughtred2019biogrid}
R.~Oughtred, C.~Stark, B.-J. Breitkreutz, J.~Rust, L.~Boucher, C.~Chang,
  N.~Kolas, L.~O’Donnell, G.~Leung, R.~McAdam, {\em et~al.}, ``{The BioGRID
  interaction database: 2019 update},'' {\em Nucleic Acids Research}, vol.~47,
  no.~D1, pp.~D529--D541, 2019.

\bibitem{kerrien2012intact}
S.~Kerrien, B.~Aranda, L.~Breuza, A.~Bridge, F.~Broackes-Carter, C.~Chen,
  M.~Duesbury, M.~Dumousseau, M.~Feuermann, U.~Hinz, {\em et~al.}, ``{The
  IntAct molecular interaction database in 2012},'' {\em Nucleic Acids
  Research}, vol.~40, no.~D1, pp.~D841--D846, 2012.

\bibitem{szklarczyk2021string}
D.~Szklarczyk, A.~L. Gable, K.~C. Nastou, D.~Lyon, R.~Kirsch, S.~Pyysalo, N.~T.
  Doncheva, M.~Legeay, T.~Fang, P.~Bork, {\em et~al.}, ``{The STRING database
  in 2021: customizable protein--protein networks, and functional
  characterization of user-uploaded gene/measurement sets},'' {\em Nucleic
  Acids Research}, vol.~49, no.~D1, pp.~D605--D612, 2021.

\bibitem{licata2012mint}
L.~Licata, L.~Briganti, D.~Peluso, L.~Perfetto, M.~Iannuccelli, E.~Galeota,
  F.~Sacco, A.~Palma, A.~P. Nardozza, E.~Santonico, {\em et~al.}, ``{MINT, the
  molecular interaction database: 2012 update},'' {\em Nucleic Acids Research},
  vol.~40, no.~D1, pp.~D857--D861, 2012.

\bibitem{peng2017protein}
X.~Peng, J.~Wang, W.~Peng, F.-X. Wu, and Y.~Pan, ``{Protein--protein
  interactions: detection, reliability assessment and applications},'' {\em
  Briefings in Bioinformatics}, vol.~18, no.~5, pp.~798--819, 2017.

\bibitem{erdHos1960evolution}
P.~Erd{\H{o}}s, A.~R{\'e}nyi, {\em et~al.}, ``{On the evolution of random
  graphs},'' {\em Publ. Math. Inst. Hung. Acad. Sci}, vol.~5, no.~1,
  pp.~17--60, 1960.

\bibitem{newman2018networks}
M.~Newman, {\em {Networks}}.
\newblock Oxford university press, 2018.

\bibitem{obayashi2022atted}
T.~Obayashi, H.~Hibara, Y.~Kagaya, Y.~Aoki, and K.~Kinoshita, ``{ATTED-II v11:
  a plant gene coexpression database using a sample balancing technique by
  subagging of principal components},'' {\em Plant and Cell Physiology},
  vol.~63, no.~6, pp.~869--881, 2022.

\bibitem{yu2010gosemsim}
G.~Yu, F.~Li, Y.~Qin, X.~Bo, Y.~Wu, and S.~Wang, ``{GOSemSim: an R package for
  measuring semantic similarity among GO terms and gene products},'' {\em
  Bioinformatics}, vol.~26, no.~7, pp.~976--978, 2010.

\bibitem{saelens2018comprehensive}
W.~Saelens, R.~Cannoodt, and Y.~Saeys, ``{A comprehensive evaluation of module
  detection methods for gene expression data},'' {\em Nature Communications},
  vol.~9, no.~1, p.~1090, 2018.

\bibitem{schubert2021fast}
E.~Schubert and P.~J. Rousseeuw, ``{Fast and eager k-medoids clustering: O (k)
  runtime improvement of the PAM, CLARA, and CLARANS algorithms},'' {\em
  Information Systems}, vol.~101, p.~101804, 2021.

\bibitem{hartigan1979algorithm}
J.~A. Hartigan and M.~A. Wong, ``{Algorithm AS 136: A k-means clustering
  algorithm},'' {\em Journal of the Royal Statistical Society. Series C
  (applied Statistics)}, vol.~28, no.~1, pp.~100--108, 1979.

\bibitem{bezdek1984fcm}
J.~C. Bezdek, R.~Ehrlich, and W.~Full, ``{FCM: The fuzzy c-means clustering
  algorithm},'' {\em Computers \& Geosciences}, vol.~10, no.~2-3, pp.~191--203,
  1984.

\bibitem{van2000graph}
S.~M. Van~Dongen, {\em {Graph clustering by flow simulation}}.
\newblock PhD thesis, 2000.

\bibitem{kaufman2009finding}
L.~Kaufman and P.~J. Rousseeuw, {\em {Finding groups in data: an introduction
  to cluster analysis}}.
\newblock John Wiley \& Sons, 2009.

\bibitem{hyvarinen1999fast}
A.~Hyvarinen, ``{Fast and robust fixed-point algorithms for independent
  component analysis},'' {\em Ieee Transactions on Neural Networks}, vol.~10,
  no.~3, pp.~626--634, 1999.

\bibitem{pedregosa2011scikit}
F.~Pedregosa, G.~Varoquaux, A.~Gramfort, V.~Michel, B.~Thirion, O.~Grisel,
  M.~Blondel, P.~Prettenhofer, R.~Weiss, V.~Dubourg, {\em et~al.},
  ``{Scikit-learn: Machine learning in Python},'' {\em the Journal of Machine
  Learning Research}, vol.~12, pp.~2825--2830, 2011.

\bibitem{yao2012independent}
F.~Yao, J.~Coquery, and K.-A. L{\^e}~Cao, ``{Independent principal component
  analysis for biologically meaningful dimension reduction of large biological
  data sets},'' {\em Bmc Bioinformatics}, vol.~13, pp.~1--15, 2012.

\bibitem{kluger2003spectral}
Y.~Kluger, R.~Basri, J.~T. Chang, and M.~Gerstein, ``{Spectral biclustering of
  microarray data: coclustering genes and conditions},'' {\em Genome Research},
  vol.~13, no.~4, pp.~703--716, 2003.

\bibitem{ihmels2002revealing}
J.~Ihmels, G.~Friedlander, S.~Bergmann, O.~Sarig, Y.~Ziv, and N.~Barkai,
  ``{Revealing modular organization in the yeast transcriptional network},''
  {\em Nature Genetics}, vol.~31, no.~4, pp.~370--377, 2002.

\bibitem{li2009qubic}
G.~Li, Q.~Ma, H.~Tang, A.~H. Paterson, and Y.~Xu, ``{QUBIC: a qualitative
  biclustering algorithm for analyses of gene expression data},'' {\em Nucleic
  Acids Research}, vol.~37, no.~15, pp.~e101--e101, 2009.

\bibitem{sun2014bi}
P.~Sun, N.~K. Speicher, R.~R{\"o}ttger, J.~Guo, and J.~Baumbach, ``{Bi-Force:
  large-scale bicluster editing and its application to gene expression data
  biclustering},'' {\em Nucleic Acids Research}, vol.~42, no.~9, pp.~e78--e78,
  2014.

\bibitem{hochreiter2010fabia}
S.~Hochreiter, U.~Bodenhofer, M.~Heusel, A.~Mayr, A.~Mitterecker, A.~Kasim,
  T.~Khamiakova, S.~Van~Sanden, D.~Lin, W.~Talloen, {\em et~al.}, ``{FABIA:
  factor analysis for bicluster acquisition},'' {\em Bioinformatics}, vol.~26,
  no.~12, pp.~1520--1527, 2010.

\bibitem{wisecaver2017global}
J.~H. Wisecaver, A.~T. Borowsky, V.~Tzin, G.~Jander, D.~J. Kliebenstein, and
  A.~Rokas, ``{A global coexpression network approach for connecting genes to
  specialized metabolic pathways in plants},'' {\em The Plant Cell}, vol.~29,
  no.~5, pp.~944--959, 2017.

\bibitem{haynes2013mapping}
B.~C. Haynes, E.~J. Maier, M.~H. Kramer, P.~I. Wang, H.~Brown, and M.~R. Brent,
  ``{Mapping functional transcription factor networks from gene expression
  data},'' {\em Genome Research}, vol.~23, no.~8, pp.~1319--1328, 2013.

\bibitem{koschutzki2008centrality}
D.~Kosch{\"u}tzki and F.~Schreiber, ``{Centrality analysis methods for
  biological networks and their application to gene regulatory networks},''
  {\em Gene Regulation and Systems Biology}, vol.~2, pp.~GRSB--S702, 2008.

\bibitem{jeong2001lethality}
H.~Jeong, S.~P. Mason, A.-L. Barab{\'a}si, and Z.~N. Oltvai, ``{Lethality and
  centrality in protein networks},'' {\em Nature}, vol.~411, no.~6833,
  pp.~41--42, 2001.

\bibitem{joy2005high}
M.~P. Joy, A.~Brock, D.~E. Ingber, and S.~Huang, ``{High-betweenness proteins
  in the yeast protein interaction network},'' {\em Journal of Biomedicine and
  Biotechnology}, vol.~2005, no.~2, p.~96, 2005.

\bibitem{wuchty2002interaction}
S.~Wuchty, ``{Interaction and domain networks of yeast},'' {\em Proteomics},
  vol.~2, no.~12, pp.~1715--1723, 2002.

\bibitem{jalili2015centiserver}
M.~Jalili, A.~Salehzadeh-Yazdi, Y.~Asgari, S.~S. Arab, M.~Yaghmaie,
  A.~Ghavamzadeh, and K.~Alimoghaddam, ``{CentiServer: a comprehensive
  resource, web-based application and R package for centrality analysis},''
  {\em Plos One}, vol.~10, no.~11, p.~e0143111, 2015.

\bibitem{ma2003connectivity}
H.-W. Ma and A.-P. Zeng, ``{The connectivity structure, giant strong component
  and centrality of metabolic networks},'' {\em Bioinformatics}, vol.~19,
  no.~11, pp.~1423--1430, 2003.

\bibitem{lareau2015differential}
C.~A. Lareau, B.~C. White, A.~L. Oberg, and B.~A. McKinney, ``{Differential
  co-expression network centrality and machine learning feature selection for
  identifying susceptibility hubs in networks with scale-free structure},''
  {\em Biodata Mining}, vol.~8, pp.~1--17, 2015.

\bibitem{ata2021recent}
S.~K. Ata, M.~Wu, Y.~Fang, L.~Ou-Yang, C.~K. Kwoh, and X.-L. Li, ``{Recent
  advances in network-based methods for disease gene prediction},'' {\em
  Briefings in Bioinformatics}, vol.~22, no.~4, p.~bbaa303, 2021.

\bibitem{kohler2008walking}
S.~K{\"o}hler, S.~Bauer, D.~Horn, and P.~N. Robinson, ``{Walking the
  interactome for prioritization of candidate disease genes},'' {\em The
  American Journal of Human Genetics}, vol.~82, no.~4, pp.~949--958, 2008.

\bibitem{brin1998pagerank}
S.~Brin, ``{The PageRank citation ranking: bringing order to the web},'' {\em
  Proceedings of Asis, 1998}, vol.~98, pp.~161--172, 1998.

\bibitem{wang2020personalized}
H.~Wang, Z.~Wei, J.~Gan, S.~Wang, and Z.~Huang, ``{Personalized pagerank to a
  target node, revisited},'' in {\em Proceedings of the 26th ACM SIGKDD
  International Conference on Knowledge Discovery \& Data Mining},
  pp.~657--667, 2020.

\bibitem{cho2015diffusion}
H.~Cho, B.~Berger, and J.~Peng, ``{Diffusion component analysis: unraveling
  functional topology in biological networks},'' in {\em International
  Conference on Research in Computational Molecular Biology}, pp.~62--64,
  Springer, 2015.

\bibitem{ideker2002discovering}
T.~Ideker, O.~Ozier, B.~Schwikowski, and A.~F. Siegel, ``{Discovering
  regulatory and signalling circuits in molecular interaction networks},'' {\em
  Bioinformatics}, vol.~18, no.~suppl\_1, pp.~S233--S240, 2002.

\bibitem{milo2002network}
R.~Milo, S.~Shen-Orr, S.~Itzkovitz, N.~Kashtan, D.~Chklovskii, and U.~Alon,
  ``{Network motifs: simple building blocks of complex networks},'' {\em
  Science}, vol.~298, no.~5594, pp.~824--827, 2002.

\bibitem{prvzulj2004modeling}
N.~Pr{\v{z}}ulj, D.~G. Corneil, and I.~Jurisica, ``{Modeling interactome:
  scale-free or geometric?},'' {\em Bioinformatics}, vol.~20, no.~18,
  pp.~3508--3515, 2004.

\bibitem{newman2004finding}
M.~E. Newman and M.~Girvan, ``{Finding and evaluating community structure in
  networks},'' {\em Physical Review E}, vol.~69, no.~2, p.~026113, 2004.

\bibitem{alsentzer2020subgraph}
E.~Alsentzer, S.~Finlayson, M.~Li, and M.~Zitnik, ``{Subgraph neural
  networks},'' {\em Advances in Neural Information Processing Systems},
  vol.~33, pp.~8017--8029, 2020.

\bibitem{adhikari2018sub2vec}
B.~Adhikari, Y.~Zhang, N.~Ramakrishnan, and B.~A. Prakash, ``{Sub2vec: Feature
  learning for subgraphs},'' in {\em Advances in Knowledge Discovery and Data
  Mining: 22nd Pacific-Asia Conference, PAKDD 2018, Melbourne, VIC, Australia,
  June 3-6, 2018, Proceedings, Part II 22}, pp.~170--182, Springer, 2018.

\bibitem{shannon2003cytoscape}
P.~Shannon, A.~Markiel, O.~Ozier, N.~S. Baliga, J.~T. Wang, D.~Ramage, N.~Amin,
  B.~Schwikowski, and T.~Ideker, ``{Cytoscape: a software environment for
  integrated models of biomolecular interaction networks},'' {\em Genome
  Research}, vol.~13, no.~11, pp.~2498--2504, 2003.

\bibitem{bastian2009gephi}
M.~Bastian, S.~Heymann, and M.~Jacomy, ``{Gephi: an open source software for
  exploring and manipulating networks},'' in {\em Proceedings of the
  international AAAI conference on web and social media}, vol.~3, pp.~361--362,
  2009.

\end{thebibliography}

\bibliographystyle{ieeetr}

\section*{Acknowledgments}
$\text{VS}^{\dagger}$ thanks Council of Scientific and Industrial Research
(CSIR), India for providing Senior Research Fellowship (SRF).
\textbf{Funding:} Authors recieved no specific funding for this research work.
\textbf{Authors Contributions:} $\text{VS}^*$ conceptualized and designed the
research framework. $\text{VS}^{\dagger}$ performed the computational
experiments. $\text{VS}^{\dagger}$ and $\text{VS}^*$ analyzed the data and
interpreted results. $\text{VS}^{\dagger}$ and $\text{VS}^*$ wrote and
finalized the manuscript.
\textbf{Competing Interests:} The authors declare that they have no conflict
%of interests.
\textbf{Data and materials availability:} NA.

\end{document}